 \documentclass[twocolumn]{aastex631}

\newcommand{\teff}[0]{\ensuremath{T_{\mathrm{eff}}}} 
\newcommand{\logg}[0]{\ensuremath{\log g}} 
\usepackage{booktabs}   
\usepackage{float}   
\usepackage{placeins} 

\begin{document}

\title{The impact of stellar rotations and binaries on the shape of upper main sequence near turn off in open cluster NGC\,6067}

\correspondingauthor{Jayanand Maurya}
\email{maurya.jayanand@gmail.com}

\correspondingauthor{Yu Zhang}
\email{zhy@xao.ac.cn}

\author[0000-0001-5119-8983]{Jayanand Maurya}
\affiliation{Xinjiang Astronomical Observatory, Chinese Academy of Sciences, No. 150, Science 1 Street, Urumqi, Xinjiang 830011, P. R. China}

\author[0000-0001-7134-2874]{Yu Zhang}
\affiliation{Xinjiang Astronomical Observatory, Chinese Academy of Sciences, No. 150, Science 1 Street, Urumqi, Xinjiang 830011, P. R. China}
\affiliation{School of Astronomy and Space Science, University of Chinese Academy of Sciences, Beijing 100049, P. R. China}

\author[0000-0001-6604-0505]{Sebastian Kamann}
\affiliation{Astrophysics Research Institute, Liverpool John Moores University, IC2 Liverpool Science Park, 146 Brownlow Hill, Liverpool L3 5RF, UK}

\author[0000-0001-5796-8010]{Hubiao Niu}
\affiliation{Xinjiang Astronomical Observatory, Chinese Academy of Sciences, No. 150, Science 1 Street, Urumqi, Xinjiang 830011, P. R. China}

\author[0000-0002-4645-6017]{Yves Fr{\'e}mat}
\affiliation{Royal Observatory of Belgium, Avenue circulaire 3, 1180 Bruxelles,
Belgium}

\author[0009-0006-1313-4675]{Kaixiang Lang}
\affiliation{Xinjiang Astronomical Observatory, Chinese Academy of Sciences, No. 150, Science 1 Street, Urumqi, Xinjiang 830011, P. R. China}

\author[0000-0001-8657-1573]{Y. C. Joshi}
\affiliation{Aryabhatta Research Institute of Observational Sciences (ARIES), Manora peak, Nainital 263002, India}

\author[0000-0002-9431-6297]{M. R. Samal}
\affiliation{Astronomy \& Astrophysics Division, Physical Research Laboratory, Ahmedabad 380009, India}

\author[0000-0001-5419-2042]{Peter De Cat}
\affiliation{Royal Observatory of Belgium, Avenue circulaire 3, 1180 Bruxelles,
Belgium}

\author[0000-0003-1845-4900]{Ali Esamdin}
\affiliation{Xinjiang Astronomical Observatory, Chinese Academy of Sciences, No. 150, Science 1 Street, Urumqi, Xinjiang 830011, P. R. China}
\affiliation{School of Astronomy and Space Science, University of Chinese Academy of Sciences, Beijing 100049, P. R. China}


\begin{abstract}
We present the analysis of the extended Main Sequence Turn-Off (eMSTO) in the open cluster NGC\,6067. We derive the projected rotational velocity, \textit{v}sin\textit{i}, of the stars belonging to the eMSTO region of the main sequence (MS) utilizing \textit{Gaia}-ESO spectra. Our results reveal a positive correlation between \textit{v}sin\textit{i} and the color of eMSTO stars, where fast-rotating stars predominantly occupy the red part of the MS while slow-rotating ones prefer a bluer side of the MS. The gravity-darkening effect might be a reason for this correlation. We find that most of the close binaries present in the eMSTO population would be slow-rotating due to the tidal-locking phenomenon. We identify four double-lined spectroscopic binaries (SB2) featuring slow-rotating companions, further supporting this tidal-locking hypothesis. However, the spatial distribution and the cumulative radial distribution indicate a higher concentration of red eMSTO stars in the cluster's central region than their bluer counterparts. This suggests that tidal locking is less likely to be the cause of the observed spread in rotation rates among eMSTO stars. Instead, we propose that star-disk interactions during the pre-main-sequence phase might have played a crucial role in spreading the rotation rates of stars, leading to the eMSTO phenomenon in NGC\,6067.

\end{abstract}



\section{Introduction} \label{sec:intro}

The presence of the eMSTO in open clusters has attracted scientific curiosity about the stellar composition of the clusters. The extended star formation over a period of 100-500 Myr has been initially attributed to the origin of the eMSTO in star clusters \citep{2009A&A...497..755M,2014ApJ...797...35G}. The second-generation stars born out of extended star formation are expected to be less populous than the first-generation stars, which does not hold true in observations \citep{2009MNRAS.394..124B}. Additionally, there has not been any detection of the extended star formation on a scale of a few hundred million years \citep{2016MNRAS.457..809C}. Considering the shortcomings of the models proposed for the extended star formation, \citet{2009MNRAS.398L..11B} proposed that the spread in rotation velocity of the stars in the upper MS can mimic the eMSTO in star clusters. The reduced gravity near the equator of stars due to rotation manifests into lower effective temperatures and luminosities for fast-rotating stars \citep{1924MNRAS..84..665V}. This gravity-darkening effect, combined with the viewing angle for the stars, may lead to a spread in color in the upper MS and, hence, can cause the eMSTO in the clusters \citep{2024MNRAS.532.1212M}. Additionally, the rotational mixing in the stars can modify the chemical composition of the stellar envelope and enhance the stellar core size, which results in lower temperatures and higher luminosities for the fast-rotating stars \citep{2003A&A...399..603P}. Many previous studies also find observational evidence that the color of the eMSTO stars in the color-magnitude diagram (CMD) is correlated with their rotational velocity \citep{2018MNRAS.480.3739B,2020MNRAS.492.2177K,2021MNRAS.502.4350S}.

All the MS stars with masses below $\sim$1.6 M$_{\odot}$ are expected to be slow-rotating because of the magnetic braking of their rotations \citep{1967ApJ...150..551K}. So, if the eMSTO in the open clusters is to be caused by the spread in rotation velocity, then the group of the eMSTO stars should be heavier than $\sim$1.6 M$_{\odot}$ to also comprise fast-rotating stars. This scenario for the lower mass limit of the eMSTO stars is also found to be true in some previous studies \citep{2020MNRAS.492.2177K,2020MNRAS.495.1978B}. However, the lower mass limit is predicted to depend on the open clusters' metallicity \citep{2019A&A...622A..66G}. Older clusters cannot host massive stars, so the lower mass limit indicates a certain age limit for the open clusters to host the eMSTO. Only clusters below 2 Gyr of age are predicted to host the eMSTO in the Large Magellanic Cloud  \citep{2019A&A...622A..66G}. The open clusters below 2 Gyr ages are found to host the eMSTO in the Milky Way Galaxy \citep{2023A&A...672A..29C}. However,  the age limit at which the eMSTO disappears shifts toward older age with increasing metallicity \citep{2019A&A...622A..66G}. 

The possible mechanisms that may produce the observed spread in rotational velocity of the eMSTO stars having similar masses are also interesting to explore. Interacting binaries may play a role in producing the spread in the rotational velocity of the eMSTO stars. \citet{2015MNRAS.453.2637D} suggested that tidal interactions in binary stars cause the initially fast-rotating stars to slow down and thus produce the spread in the rotational velocity of the stars. In such a scenario, tidally locked binaries would populate the blue part of the eMSTO and are expected to be preferentially located in the central region of the cluster. However, \citet{2024MNRAS.532.1212M} found that the stars in the blue part of the eMSTO were preferentially located in the outer region of the cluster NGC 2355. Similar spatial distribution patterns for the blue and the red parts of the eMSTO populations were also reported in previous studies \citep{2023MNRAS.518.1505K,2024A&A...689A.162N}. If tidal interactions in binaries were the main rotation braking mechanism, blue eMSTO stars should exhibit a higher binary fraction than red eMSTO stars. This is because blue eMSTO stars tend to rotate more slowly, which aligns with the idea that binary interactions slow down stellar rotation. However, similar binary fractions have been found across the eMSTO region in the CMD for a few instances, which appear to be at odds with the proposed binary interaction mechanism for the origin of the spread in rotational velocity of the eMSTO stars \citep{2020MNRAS.492.2177K,2021MNRAS.508.2302K}. 

An alternative model to the interacting binary model explaining the spread in rotational velocity of the eMSTO in star clusters is proposed by \citet{2020MNRAS.495.1978B}. In this model, the observed spread in the rotational velocity of the eMSTO stars is attributed to the spread in the rotational velocity distribution of the stars during their pre-main-sequence (PMS) phase due to star–disk interaction (SDI) phenomena. The stars lose angular momentum due to the coupling of their circumstellar disk. The stars that lose their circumstellar disk early become free to spin up due to the star's contraction to the MS. Thus, the stars retaining their circumstellar disk for a longer period rotate slowly compared to the stars losing the disk earlier. The spread in rotation velocity during the PMS phase persists up to the age of $\sim$1.5 Gyr for stars on the MS. The SDI causing the spread in rotation rates of the stars is suggested to be the reason for the origin of the eMSTO in open cluster NGC 2355 \citep{2024MNRAS.532.1212M}.

\citet{2022NatAs...6..480W} suggest that the binary mergers may produce slow-rotating blue eMSTO stars in star clusters. According to this theory, a binary merger will rejuvenate the stellar core with more hydrogen fuel, which causes the star to appear bluer than coeval single stars of the same mass. The merger product star rotates slowly due to the loss of angular momentum during the thermal expansion phase post-coalescence and internal restructuring.  

The eMSTO in star clusters has been mostly attributed to the spread in rotational velocity of the stars. However, other mechanisms have also been suggested to contribute to the origin of the eMSTO in the star clusters. The dust-like extinction from an excretion disk of the fast-rotating star can cause the star to appear redder than its non-rotating counterparts, which may produce the eMSTO \citep{2023MNRAS.521.4462D}. The differential reddening across the clusters may also resemble the eMSTO \citep{2021A&A...656A.149A}. Similarly, binary stars with different mass ratios can create a redder spread in color on the MS compared to the single stars resembling the eMSTO \citep{2022MNRAS.512.3992C}.       

In this paper, we present the study of the eMSTO in the open cluster NGC\,6067. This cluster is known to host Be stars and SB2s, which are interesting objects to understand the role of stellar rotation and tidally interacting binaries in the origin of the eMSTO in open clusters \citep{2017MNRAS.469.1330A}. NGC\,6067, aged 126 Myr, is among the relatively young open clusters for hosting the eMSTO \citep{2020A&A...640A...1C}. \citet{2017MNRAS.469.1330A} estimated the mean radial velocity of the cluster to be $-$39.5$\pm$0.9 km s$^{-1}$ using high-resolution (R = 48,000) spectra of 45 member stars. 
\section{Data}
\subsection{Photometric}
We utilized \textit{Gaia} Data Release 3 (DR3) astrometric and photometric data to identify member stars and further analyze the cluster NGC\,6067 \citep{2016A&A...595A...1G,2023A&A...674A...1G}. \textit{Gaia} DR3 astrometric data is especially useful for the study of the eMSTO phenomenon as it provides a very high accuracy of $\sim$0.03 mas yr$^{-1}$ in the proper motions and a similar accuracy of $\sim$0.03 mas in parallax for bright stars with G$<$15 mag \citep{2023A&A...674A...1G}. The astrometric data from \textit{Gaia} DR3 has median uncertainties of 0.07 and 0.5 mas yr$^{-1}$ in proper motions and 0.07 and 0.5 mas in parallax for stars up to G=17 and G=20 mag, respectively. The \textit{Gaia} DR3 provides good accuracy for the photometric data, too. It has uncertainties of 0.001, 0.012, and 0.006 mag in the G, G$_{\rm BP}$, and G$_{\rm RP}$ bands for bright stars up to 17 mag. \textit{Gaia} DR3 has relatively high photometric uncertainties of 0.006, 0.108, and 0.052 mag for stars with G = 20 mag.  

\subsection{Spectroscopic}
We used the spectroscopic data from the European Southern Observatory (ESO) archives observed under program IDs 188.B-3002(M) and 095.C-0818(A). The spectra were observed from the Ultraviolet and Visual Echelle Spectrograph (UVES) and GIRAFFE spectrographs mounted on the ESO Very Large Telescope. The spectral resolutions ($\frac{\lambda}{\Delta \lambda}$) for UVES and GIRAFFE are $\sim$51000 and 29000, respectively. These spectra typically have a signal-to-noise ratio above $\sim$50. The spectra from UVES have a wavelength range from 420 to 620 nm. The spectra from GIRAFFE are in the wavelength ranges 385-405 nm and 515-535 nm. In the \textit{Gaia}-ESO archives, we found spectra for 41 stars, including four SB2 and two Be stars, belonging to the upper MS of NGC\,6067. Seven of these 41 stars have high-resolution UVES spectra; the remaining have GIRAFFE spectra. We also utilized \textit{Gaia}-ESO spectra for two blue straggler stars (BSS) and 14 red giant branch stars (RGB) to estimate their rotational and radial velocity. Only one of the two BSS stars has UVES spectra. All the spectra of giant stars were taken with the UVES spectrograph. In general, only one or two spectra are available in the \textit{Gaia}-ESO archives for each of these upper MS, BSS, and RGB stars.   

\section{Physical properties}
\subsection{Membership and physical parameters}
\begin{figure}
	\includegraphics[width=8.5 cm]{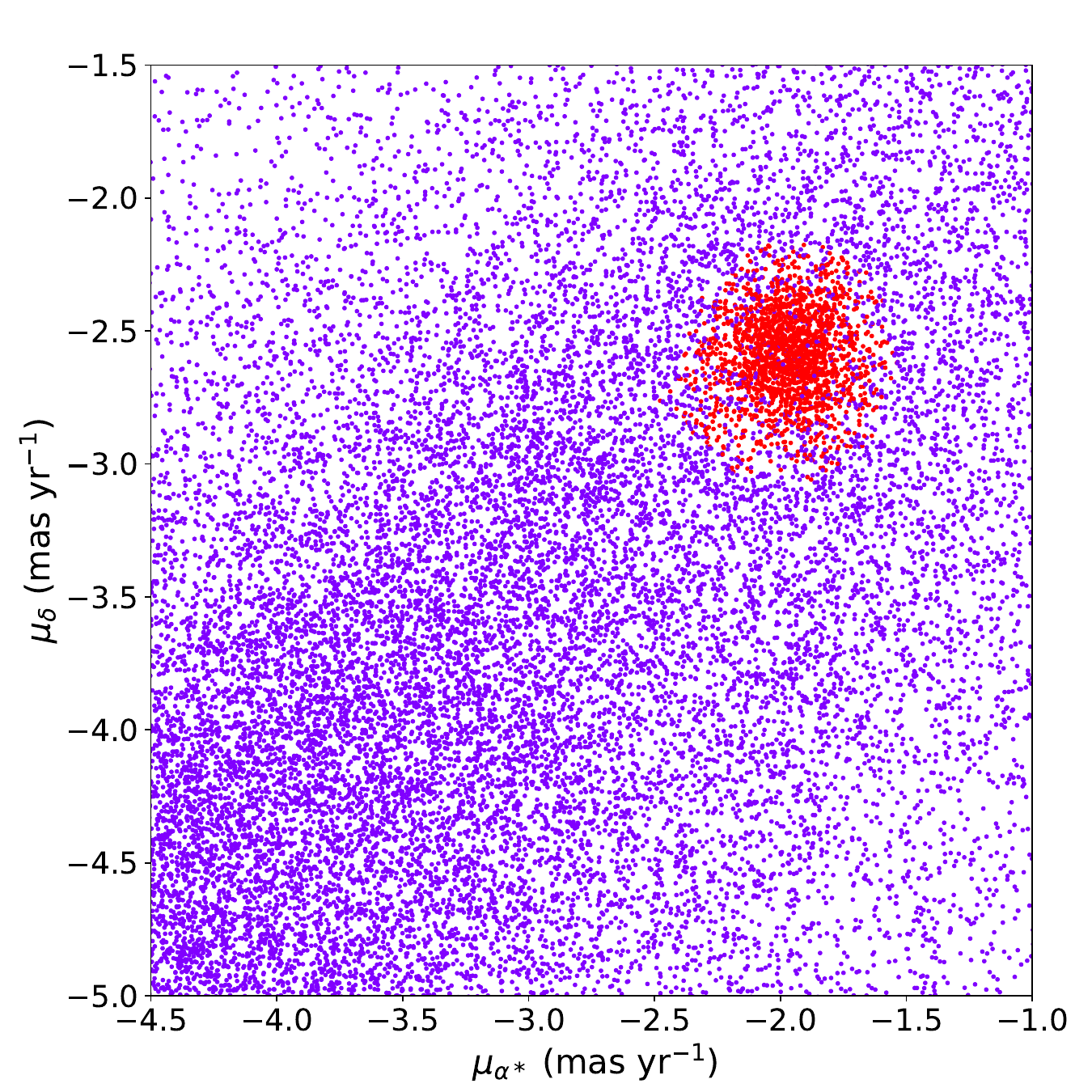}
    \caption{The cluster over-density in the proper motions space. The red points represent potential cluster member stars. Blue points denote the field stars.}
    \label{over-density}
\end{figure}

We identified the member stars of NGC\,6067 using the Hierarchical Density-Based Spatial Clustering of Applications with Noise (HDBSCAN) algorithm \citep{2017JOSS....2..205M}. We used \textit{Gaia} DR3 proper motions and parallaxes to identify clustering and obtain the membership probability of stars \citep{2016A&A...595A...1G,2023A&A...674A...1G}. We used proper motions ($\mu_{\alpha *}$, $\mu_{\delta}$) and parallax ($\varpi$) as input parameters for the over-density identification through the HDBSCAN. To avoid the inclusion of bad data points, we included only those stars with parallax uncertainty below 0.25 mas in our sample for over-density identification. We chose minimum cluster size, minimum samples, and maximum cluster size to be 400, 400, and 1800 for over-density detection. The resulting over-density is shown in Figure~\ref{over-density}. Stars having a membership probability greater than 70$\%$ were considered as member stars. We then plotted the cluster's CMD and removed nine outlier stars, which were found to be significantly deviating from the MS in visual inspection of the CMD. These stars have G magnitude greater than 14.6 mag. Four out of the nine outliers are fainter than 18 mag in the G band. Finally, we obtained 944 member stars in the cluster NGC\,6067. Out of the 944 member stars identified by us, 867 and 633 are common with those listed by \citet{2023A&A...673A.114H} and \citet{2020A&A...640A...1C}, respectively. \textit{Gaia}-DR3 offers proper motions and parallaxes for 10$\%$ more sources than \textit{Gaia}-DR2, with significantly improved precision \citep{2021A&A...649A...2L,2021A&A...649A...5F}. The increased number of sources with astrometric solutions and better precision likely accounts for the smaller overlap of member stars with the \citet{2020A&A...640A...1C} catalog, which is based on \textit{Gaia}-DR2 data. Additionally, \citet{2020A&A...640A...1C} includes stars only up to G = 17.8 mag for NGC\,6067, while our sample extends to G = 18.6 mag, which may also explain the lower number of common members.

\begin{figure*}
	\includegraphics[width= 17.5 cm]{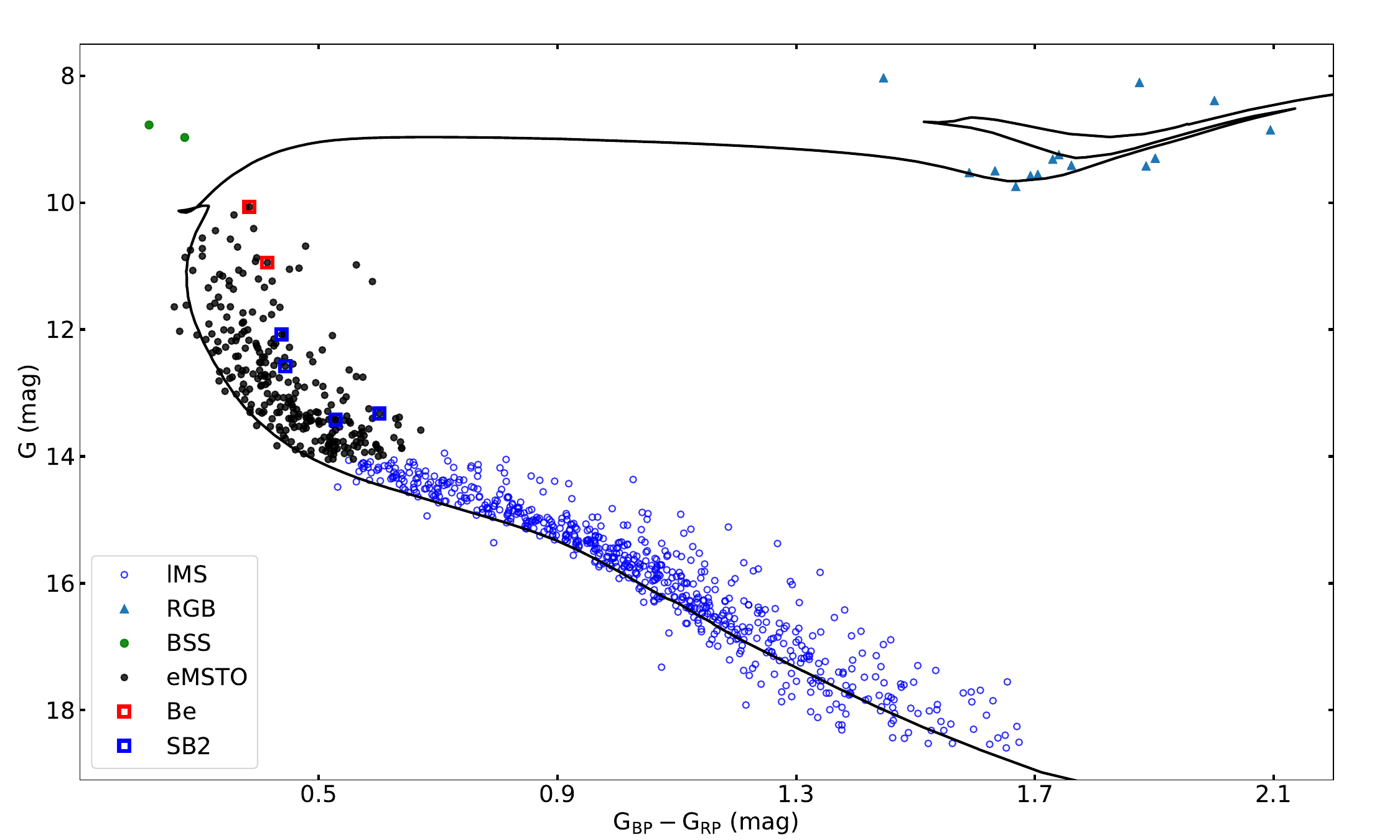}
    \caption{The color-magnitude diagram of the cluster NGC\,6067. The stars belonging to the eMSTO are shown by black points. The eMSTO stars also include Be stars enclosed by red squares and SB2 stars enclosed by blue squares. The RGB, BSS, and lMS stars are also shown by the markers given in the legend of the figure. The black continuous curve shows the best-fit \citet{2017ApJ...835...77M} isochrone corresponding to the logarithmic age of 7.96 years.}
    \label{cmd_binary}
\end{figure*}

The mean parallax of the member stars is calculated to be 0.472$\pm$0.045 mas, which corresponds to a distance of 1960$\pm$89 pc after applying the Global systematic offset of $-$0.039 mas in the parallax \citep{2021A&A...654A..20G}. The mean parallax and distance obtained agree well with their values as 0.471$\pm$0.038 mas and 1957 pc provided by \citet{2023A&A...673A.114H}. We fitted an extinction-corrected isochrone to the CMD constructed from the identified member stars to estimate the cluster's age. We took the extinction value to be A$_{V}$ = 0.99 mag as provided by \citet{2023A&A...673A.114H}. The best fit was obtained for an isochrone corresponding to a log(age) of 7.96 years, which corresponds to a cluster age of 91 Myr. We used metallicity Z = 0.02 for the isochrones which was calculated from the [Fe/H] abundance of $-$0.11$\pm$0.02 dex as estimated in Section~\ref{spectro_param}. The physical parameters of NGC\,6067 are summarized in Table~\ref{cluster_param}. The extinction-corrected \citet{2017ApJ...835...77M} isochrone fitted on the CMD of NGC\,6067 is shown in Figure~\ref{cmd_binary}. As clearly visible in the figure, we detected a broadened upper MS resembling an eMSTO. We considered the MS stars having G band magnitudes from 9.85 to 14.05 mag and G$_{\rm BP}-$G$_{\rm RP}$ color between 0.20 to 0.70 mag to be the eMSTO stars. This way, we found 280 eMSTO stars in the CMD of NGC\,6067, as shown by the black points in
Figure~\ref{cmd_binary}. The masses of the eMSTO stars are estimated to be in the range 2.1-5.3 M$_{\odot}$ through isochrone fitting on the CMD of the cluster. The MS stars fainter than the eMSTO stars and lying in the lower MS are termed as lMS stars.   
\begin{table}
\caption{Physical parameters of NGC\,6067 obtained in the present study. We have also provided values of these parameters from the literature for comparison. References: 1- present study; 2- \citet{2020A&A...640A...1C}; 3- \citet{2023A&A...673A.114H}; 4- \citet{2022AJ....163..195R}; 5- \citet{2017MNRAS.469.1330A}}
\begin{center} \label{cluster_param}
\begin{tabular}{cccc}  
\hline
log (age) & mean parallax& Distance& [Fe/H] \\
(yr)&         (mas)&  pc     &   (dex) \\
\hline     
 7.96 (1)& 0.472$\pm$0.045 (1)& 1960$\pm$89 (1)& $-$0.11$\pm$0.02 (1)\\
 8.10 (2)& 0.471$\pm$0.038 (3)& 1957 (3)& +0.03$\pm$0.27 (4) \\
        -&                   -&        -& +0.19$\pm$0.05 (5) \\
\hline
\end{tabular}
\end{center}
\end{table}
\subsection{Differential reddening}
\begin{figure}
	\includegraphics[width=8.5 cm]{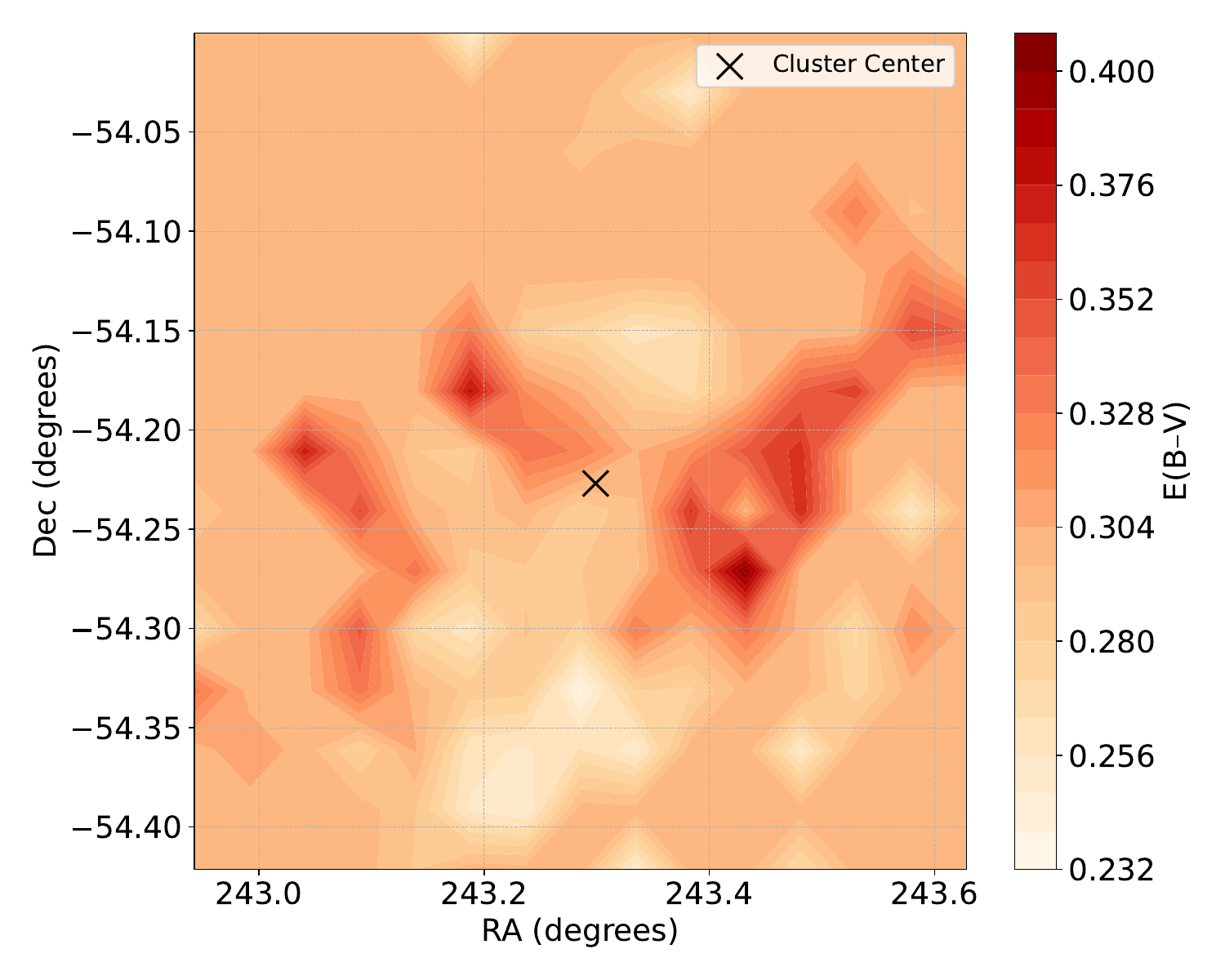}
    \caption{2D color-map of extinction for the eMSTO stars in NGC\,6067 color-coded by A$_{V}$ values taken from \citet{2024A&A...691A..98K}.}
    \label{2D_AV}
\end{figure}

The differential reddening in an open cluster region may also contribute to the G$_{\rm BP}-$G$_{\rm RP}$ color spread in the upper MS of the cluster. We utilized the line-of-sight extinction, A$_{V}$, in the V band provided by \citet{2024A&A...691A..98K} to investigate the extinction across NGC\,6067. \citet{2024A&A...691A..98K} estimated A$_{V}$ values with a median uncertainty of 0.20 mag through the gradient-boosted random-forest regressor (xgboost) algorithm using \textit{Gaia} DR3 multi-band photometry, astrometry, and most notably XP spectra. We have shown the 2D reddening, E(B$-$V), map of the eMSTO stars of NGC\,6067 in Figure~\ref{2D_AV}. We calculated the E(B$-$V) from A$_{V}$ values using the relation E(B$-$V) = A$_{V}$/3.1, based on the extinction law with R$_{V}$ = 3.1 \citep{1989ApJ...345..245C}. The 2D reddening map for the eMSTO stars is shown in Figure~\ref{2D_AV}. The reddening map reveals a mild differential reddening with E(B$-$V) varying from 0.23 to 0.40 mag for the eMSTO stars across the NGC\,6067 cluster, considering the inherent median uncertainty of 0.20 mag in A$_{V}$ values taken from \citet{2024A&A...691A..98K}. \citet{2016RMxAA..52..223T} reported similar color excesses but slightly shifted towards higher values for NGC\,6067, ranging from  E(B-V) = 0.30 to 0.48 mag. 

\begin{figure*}
	\includegraphics[width=17.5 cm]{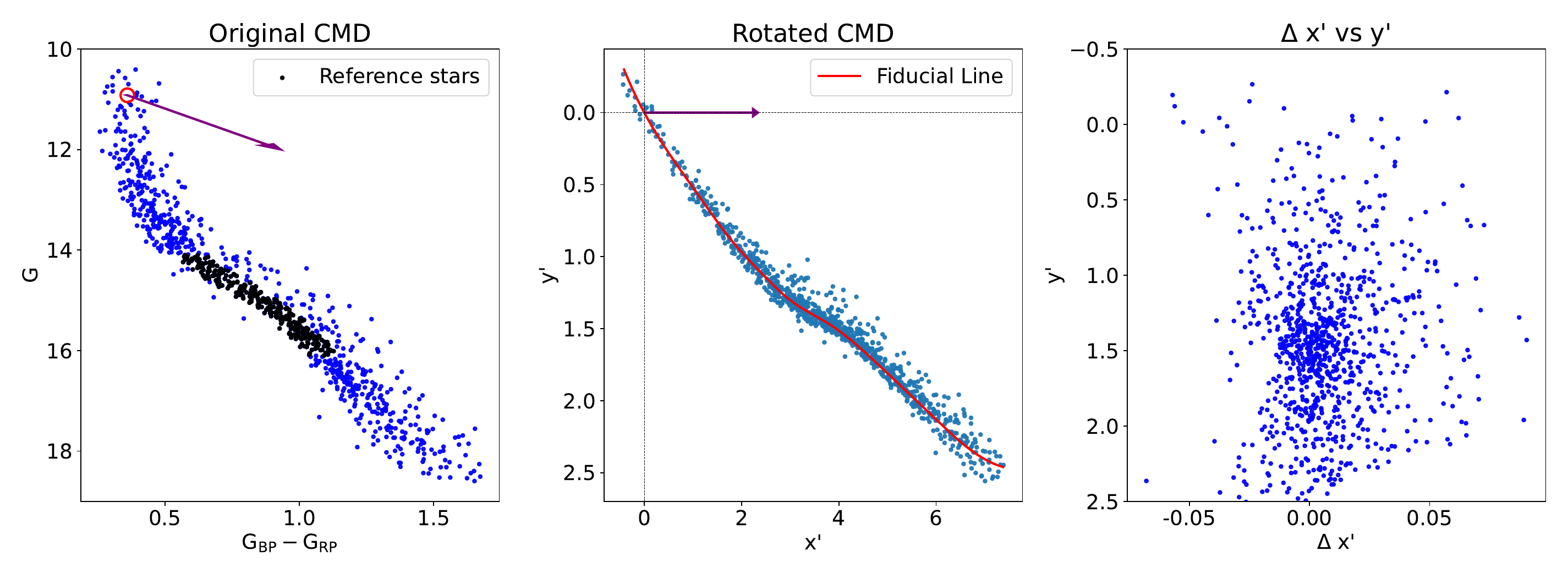}
    \caption{Left panel: observed color-magnitude diagram of NGC\,6067 where the reddening vector is shown by an arrow. Middle panel: CMD in the rotated frame which is considered as a new reference frame for the differential reddening measurements. The red curve represent the fiducial line in the new reference frame. The reddening vector which becomes parallel to the x'-axis in the new reference frame is shown by an arrow. Right panel: a plot of y' versus $\Delta$x' estimated in the new reference frame.}
    \label{RedVec}
\end{figure*}

To correct for differential reddening, we followed the method outlined by \citet{2012A&A...540A..16M} and generated a differential reddening-corrected MS for NGC\,6067. Here, we briefly outline the procedure. We initially arbitrarily chose a point (O) near MS turn-off in the CMD as shown by the red circle in left panel of Figure~\ref{RedVec}. Next, we translated the CMD so that the point corresponding to O becomes origin of the new reference frame. Lastly, we rotated the CMD counterclockwise by an angle $\theta$ to align the abscissa with the reddening vector of NGC\,6067 to easily estimate reddening differences as shown in middle panel of Figure~\ref{RedVec}. The rotation angle $\theta$ can be calculated using the following formula:  

\[
\theta = \arctan\left(\frac{\mathrm{A}_\mathrm{G}}{\mathrm{E}(\mathrm{G}_{\mathrm{BP}} - \mathrm{G}_{\mathrm{RP}})}\right)
\]

\begin{figure*}
	\includegraphics[width=17.5 cm]{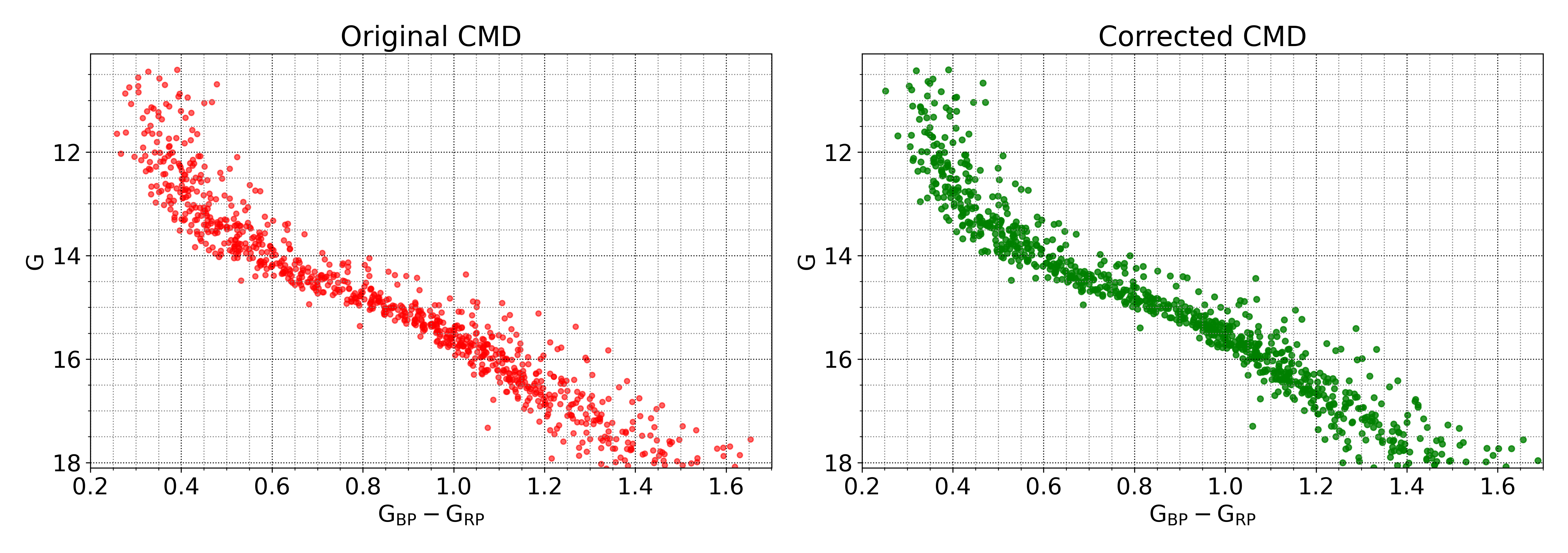}
    \caption{The comparison between original CMD (left panel) and differential reddening corrected CMD (right panel) of NGC\,6067 cluster. }
    \label{corrected_CMD}
\end{figure*}

For each MS star in the rotated CMD, we calculated the distance from the fiducial line along the reddening vector, which gives the residual color ($\Delta$x') for the particular star. A plot of the y' versus $\Delta$x' is shown in right panel of Figure~\ref{RedVec}. We selected MS stars with G-magnitudes between 14.1 and 16.1 mag as reference stars, as depicted by black points in the left panel of Figure~\ref{RedVec}. We selected only those stars on the MS in this magnitude range as the reference stars whose distance from the fiducial line was within 80 percentile to avoid inclusion of the stars with large observational errors or binary companion contamination. For each MS star, we calculated median of the residual color of 50 nearest reference stars. This median residual color was taken as the local differential reddening estimate for each star. We then subtracted these median residual color from the residual color of the individual stars to obtain their differential reddening-corrected colors. A comparison between the original MS and the differential reddening-corrected MS in the CMD is shown in Figure~\ref{corrected_CMD}. While the corrected MS appears slightly narrower, the eMSTO is still present, suggesting that the eMSTO in MS of NGC\,6067 is intrinsic and not entirely caused by differential reddening.

\section{Spectroscopic properties of the stars}\label{spectro_param}
We estimated the radial velocity (RV) and the projected rotational velocity, \textit{v}sin\textit{i},  of the stars using spectra from the \textit{Gaia}-ESO archives. The \textit{iSpec} software package was used to derive these velocities \citep{2014A&A...569A.111B,2019MNRAS.486.2075B}. We used the atomic lines from the Vienna Atomic Line Database (VALD) provided by \citet{1999A&AS..138..119K,2000BaltA...9..590K} to estimate the RV values. The velocity profile of the stars for the RV calculation is generated through a cross-match correlation algorithm. The \textit{iSpec} package integrates many atmospheric models and radial transfer codes together to generate synthetic spectra. The \textit{v}sin\textit{i} values for the stars were estimated by fitting the synthetic spectra on the observed spectra of the stars through the $\chi^{2}$ minimization process. We used the \textit{ATLAS9} atmospheric models with the \textit{SPECTRUM} radial transfer codes to generate the synthetic spectra \citep{2005MSAIS...8...14K,1994AJ....107..742G}. We adopted the reference solar abundances from \citet{2007SSRv..130..105G}. The microturbulence velocity, $\xi$, for the stars except RGB stars was taken as a fixed value of 2 km s$^{-1}$ \citep{2011A&A...528A..44M}. We took advantage of the high-resolution UVES data available for RGB stars to estimate the $\xi$ values for each of them iteratively, as discussed in Section~\ref{RGB}. Following the described process, we estimated \textit{v}sin\textit{i} values for 35 out of 41 eMSTO stars possessing spectra, 2 BSS stars, and 14 RGB stars. The \textit{v}sin\textit{i} values for the remaining 6 eMSTO stars possessing spectra, including two Be stars and four SB2 stars, were estimated using the GIRFIT code \citep{2006A&A...451.1053F} as described in Sections~\ref{Be_sect} and~\ref{SB2_sect}. The obtained \textit{v}sin\textit{i} values with other physical parameters of the eMSTO and RGB stars are given in Tables~\ref{eMSTO_param} and ~\ref{RGB_param}, respectively. 

The eMSTO stars in NGC\,6067 have a wide range of \textit{v}sin\textit{i} values from $\sim$9 to 336 km s$^{-1}$. The eMSTO stars in NGC\,6067 have masses in the range of $\sim$2.1-5.3 M$_{\odot}$. The stars with similar masses on the MS should be fast-rotating in the absence of magnetic braking. However, we found the spread in the rotation rates of the eMSTO stars of similar masses. The possible causes for the spread in the rotation rates of the stars, particularly of the eMSTO stars, are discussed in Section~\ref{discuss}. From \citet{2022A&A...666A.121R} we utilized \textit{v}sin\textit{i} values for 147 lMS stars in NGC\,6067, though their data exclude eMSTO and RGB stars of the NGC\,6067 cluster. The lMS stars were mostly slow-rotating stars with an average \textit{v}sin\textit{i} value of 103 km s$^{-1}$.  
\begin{table*}
\caption{The list of the eMSTO stars of the cluster NGC\,6067 whose \textit{v}sin\textit{i} values are estimated in the present study. The ID, right ascension, declination, \textit{v}sin\textit{i}, G band magnitude,  G$_{\rm BP}-$G$_{\rm RP}$ color, extinction in V band, and mass of the eMSTO stars are given in columns 1, 2, 3, 4, 5, 6, 7, and 8, respectively.} \label{eMSTO_param}
\begin{tabular}{ccccccccccccccc}  
\hline
ID  &    RA (J2000)&   Dec. (J2000)& \textit{v}sin\textit{i}&        G& G$_{\rm BP}-$G$_{\rm RP}$& A$_{V}$& Mass \\
     & (deg)& (deg)& (km s$^{-1}$)& (mag)& (mag)& (mag)& (M$_{\odot}$) \\
\hline     
   5& 243.41907& $-$54.12207&  83$\pm$25& 14.0237& 0.4831&             -& 2.100\\
  21& 243.45750& $-$54.05466&  11$\pm$2 & 13.6478&  0.569& 0.91$\pm$0.16& 2.376\\
  42& 243.34326& $-$54.02435& 167$\pm$40& 12.6672&  0.334&             -& 3.243\\
 158& 243.36082& $-$54.23372&  123$\pm$7& 10.6857& 0.4779& 1.03$\pm$0.30& 4.969\\
 185& 243.42523& $-$54.20733& 204$\pm$34& 11.335 & 0.4093& 1.06$\pm$0.40& 4.525\\
 198& 243.44811& $-$54.17712& 221$\pm$30& 10.9265& 0.3936& 0.98$\pm$0.35& 4.826\\
 229& 243.19800&	$-$54.36750&   22$\pm$1& 11.6377& 0.3183& 0.84$\pm$0.20& 4.257\\
 252& 243.21793& $-$54.31174& 267$\pm$30&  11.643& 0.3357& 0.87$\pm$0.25& 4.252\\
 253& 243.18717& $-$54.31861& 192$\pm$28& 10.0726& 0.3783& 0.84$\pm$0.32& 5.287\\
 317& 243.02956& $-$54.31861&  24$\pm$8& 12.1752& 0.3607&             -& 3.731\\
 355& 243.30353& $-$54.28999& 155$\pm$15& 10.0688& 0.3839& 0.68$\pm$0.34& 5.290\\ 
 358& 243.27002& $-$54.29227&  15$\pm$11& 14.0485& 0.5158&             -& 2.083\\
 374& 243.22888& $-$54.27893& 247$\pm$25&  12.157& 0.3114&             -& 3.749\\
 377& 243.25697& $-$54.28682& 159$\pm$39& 12.8615& 0.3744& 0.80$\pm$0.17& 3.061\\\
 396& 243.30973& $-$54.28621&  132$\pm$5& 10.5764& 0.3517& 0.89$\pm$0.24& 5.036\\ 
 402& 243.31943& $-$54.27146& 295$\pm$26& 11.0653& 0.3665& 0.80$\pm$0.26& 4.735\\
 403& 243.31142& $-$54.27231&   25$\pm$5& 10.7236& 0.3051&             -& 4.950\\ 
 420& 243.31949& $-$54.25319& 172$\pm$32& 11.1322& 0.3337& 0.82$\pm$0.32& 4.682\\ 
 438& 243.31659& $-$54.24563& 298$\pm$58& 12.4412&  0.443& 0.86$\pm$0.16& 3.466\\
 459& 243.28276& $-$54.24951&   53$\pm$4& 10.8629& 0.2764&             -& 4.867\\
 558& 243.20743& $-$54.22533& 336$\pm$71& 11.5725& 0.4241& 0.94$\pm$0.28& 4.314\\
 584& 243.20327& $-$54.20562& 219$\pm$31& 11.2306& 0.3496& 0.85$\pm$0.31& 4.606\\
 594& 243.21838& $-$54.18101& 227$\pm$14& 10.6999&  0.364& 0.79$\pm$0.27& 4.962\\
 621& 243.36841& $-$54.21187& 184$\pm$40& 12.2131& 0.4196& 1.00$\pm$0.21& 3.692\\
 625& 243.33303& $-$54.20987& 300$\pm$43& 11.7253& 0.3888& 0.92$\pm$0.20& 4.174\\
 626& 243.32592& $-$54.21167& 197$\pm$25& 10.9451& 0.4137& 0.96$\pm$0.34& 4.815\\
 656& 243.30853& $-$54.19698& 237$\pm$31& 11.2045& 0.3993& 0.98$\pm$0.29& 4.627\\
 662& 243.28178& $-$54.18211& 206$\pm$22& 12.0702& 0.3715&             -& 3.842\\
 670& 243.32523& $-$54.16780& 262$\pm$15& 10.1941& 0.3577&             -& 5.213\\
 702& 243.39956& $-$54.15892& 154$\pm$24& 10.8433& 0.3055& 0.83$\pm$0.36& 4.888\\ 
 723& 243.59273& $-$54.24999& 273$\pm$39& 10.4091& 0.3906& 0.80$\pm$0.28& 5.117\\ 
 741& 243.26174& $-$54.18064& 326$\pm$20& 11.0325& 0.4674& 1.01$\pm$0.35& 4.756\\ 
 767& 243.23453& $-$54.16001& 204$\pm$10& 10.4417& 0.3272& 0.89$\pm$0.31& 5.102\\
 779& 243.30219& $-$54.14578& 306$\pm$24& 11.0685&  0.289& 0.62$\pm$0.29& 4.760\\
 797& 243.26709& $-$54.07170&    9$\pm$4& 12.7096& 0.3697& 0.91$\pm$0.22& 3.202\\
 807& 243.35750& $-$54.33215&   13$\pm$0& 11.3071& 0.3499&             -& 4.548\\  
 814& 243.31616& $-$54.33117& 154$\pm$35&  10.748& 0.2852& 0.81$\pm$0.29& 4.941\\ 
\hline
\end{tabular}
\end{table*}

\subsection{Red giant stars}\label{RGB}

We found 14 RGB stars in the cluster NGC\,6067. We utilized high-resolution (R$\sim$51,000) spectra from the UVES spectrograph to estimate the atmospheric parameters (\teff, \logg, and [Fe/H]), \textit{v}sin\textit{i}, and RV of all RGB stars. First, we estimated the RV of the RGB stars using the process discussed earlier. The estimated RV values for the RGB stars are given in Table~\ref{RGB_param}. The mean RV value for the RGB stars, excluding the star with ID 441, is found to be $-$38.78$\pm$0.02 km s$^{-1}$ with a standard deviation of 1.56 km s$^{-1}$. The star with ID 441 has an RV value of $-$45.8$\pm$0.3, which may be hosting a binary companion. After RV estimation of the individual stars, we estimated their microturbulence velocity, $\xi$, while simultaneously varying the \teff, \logg, [M/H], \textit{v}sin\textit{i}, and $\xi$. We found mean $\xi$  value as 2.74$\pm$0.03 km s$^{-1}$ with a standard deviation of 0.33 km s$^{-1}$. We then estimated  \teff, \logg, and \textit{v}sin\textit{i} values for each star by fixing $\xi$ to the obtained values and taking [M/H] as a free parameter. The estimated \teff, \logg, and \textit{v}sin\textit{i} values for the RGB stars are given in Table~\ref{RGB_param}. 

\begin{figure*}
	\includegraphics[width= 17.5 cm]{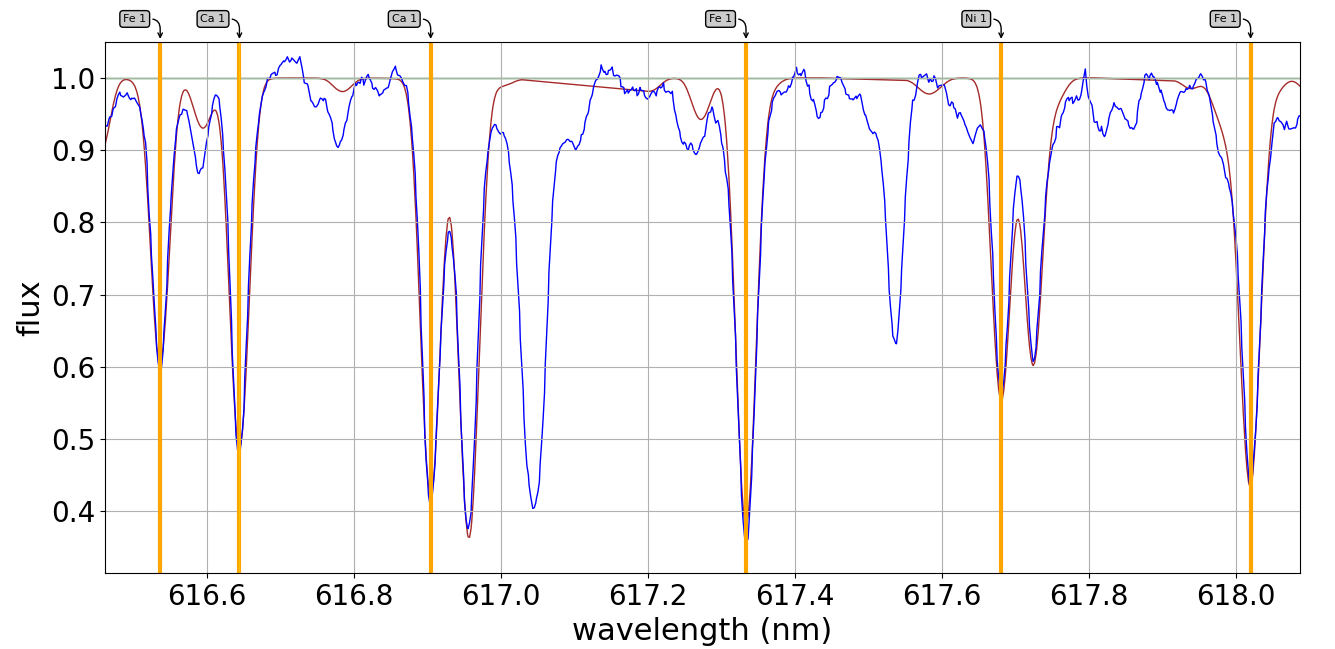}
    \caption{The observed (blue lines) spectrum and best-fit synthetic spectral lines (red lines) for the star with ID 269. The synthetic spectrum was fitted only in segments of 0.1 nm surrounding the selected lines (yellow lines) labeled by corresponding element names on the top of the plot. The best-fit synthetic spectral lines correspond to [Fe/H] = $-$0.14$\pm$0.04 dex.}
    \label{RGB_stars}
\end{figure*}

The slow rotation of the RGB stars makes them suitable candidates to identify the spectral lines corresponding to the elements present. The RGB stars have been useful in determining the [Fe/H] abundances of the open clusters as they are also brighter than the MS stars in an open cluster and exhibit sharper spectral features \citep{2013MNRAS.431.3338R}. Therefore, we estimated the [Fe/H] abundances of the RGB stars belonging to NGC\,6067. The [Fe/H] abundance of these stars was estimated through the synthetic spectral fitting technique using GES atomic line lists provided by the \textit{iSpec} package. We removed the blended lines from the line list to estimate the metallicity accurately. We also removed the lines that appeared asymmetric from the line list. We fitted synthetic spectral lines on the observed spectra in segments of 0.1 nm around the selected lines while keeping [M/H] and [Fe/H] as free parameters. The spectrum-fitting was achieved through the $\chi^{2}$ minimization technique. The best-fit synthetic spectrum overplotted on the observed spectrum for a star with ID 269 is illustrated in Figure~\ref{RGB_stars}. The [Fe/H] values of stars corresponding to the best-fit synthetic spectra are reported in Table~\ref{RGB_param}. The mean [Fe/H] for the RGB stars of NGC\,6067 is found to be $-$0.11$\pm$0.02 dex with a standard deviation of 0.09 dex. We considered the mean [Fe/H] value of $-$0.11$\pm$0.02 dex as the [Fe/H] value for the open cluster NGC\,6067. The [Fe/H] values of NGC\,6067 in the previous studies are very diverse \citep{2017MNRAS.469.1330A,2022AJ....163..195R}. The obtained mean [Fe/H] value reported here is based on a relatively larger sample of stars with more consistent values for the individual RGB stars. 
\begin{table*}
\caption{The estimated physical parameters of the RGB stars in the cluster NGC\,6067. The ID, right ascension, declination, G, G$_{\rm BP}-$G$_{\rm RP}$, radial velocity, effective temperature, surface gravity, projected rotational velocity, and [Fe/H] abundance of the RGB stars are given in columns 1, 2, 3, 4, 5, 6, 7, 8, 9, and 10, respectively.}\label{RGB_param}
\FloatBarrier
\begin{center}
\begin{tabular}{cccccccccc}  
\hline
ID  &    RA (J2000)&   Dec. (J2000)&  G& G$_{\rm BP}-$G$_{\rm RP}$& RV& T$_{\rm eff}$&  \logg& \textit{v}sin\textit{i}& [Fe/H] \\
     & (deg)& (deg)& (mag)&  (mag)&  (km s$^{-1}$)& (K)& (cm s$^{-1}$)& (km s$^{-1}$)& (dex) \\
\hline     
 216& 243.099030& $-$54.418671& 9.528& 1.590& $-$38.28$\pm$0.14& 4664.93$\pm$105.61& 1.89$\pm$0.39& 13.6$\pm$0.5& $-$0.14$\pm$0.08\\
 269& 243.064087& $-$54.322052& 9.247& 1.740& $-$39.01$\pm$0.05& 4539.85$\pm$41.44& 1.32$\pm$0.20& 7.9$\pm$0.2& $-$0.14$\pm$0.04\\ 
 286& 243.165848& $-$54.294777& 9.416& 1.761& $-$34.25$\pm$0.05& 4375.65$\pm$34.85& 1.36$\pm$0.20& 5.9$\pm$0.3& $-$0.16$\pm$0.08\\ 
 441& 243.322479& $-$54.234898& 8.036& 1.446& $-$45.80$\pm$0.29& 5793.07$\pm$131.01& 1.59$\pm$0.27& 19.5$\pm$0.7& $-$0.16$\pm$0.09\\
 457& 243.269348& $-$54.240891& 8.858& 2.095& $-$40.98$\pm$0.06& 4135.44$\pm$40.72& 1.00$\pm$0.17& 5.4$\pm$0.3& $-$0.29$\pm$0.08\\
 497& 243.266754& $-$54.205204& 8.395& 2.001& $-$40.46$\pm$0.06& 4249.77$\pm$28.96& 1.20$\pm$0.30& 5.8$\pm$0.3& $-$0.14$\pm$0.09\\
 553& 243.237076& $-$54.220432& 8.109& 1.875& $-$39.90$\pm$0.07& 4358.43$\pm$36.43& 0.84$\pm$0.17& 10.0$\pm$0.2& $-$0.16$\pm$0.04\\
 590& 243.197327& $-$54.183475& 9.578& 1.693& $-$39.48$\pm$0.09& 4757.37$\pm$51.52& 1.45$\pm$0.16& 10.2$\pm$0.2& $-$0.09$\pm$0.03\\
 602& 243.348312& $-$54.228546& 9.426& 1.886& $-$38.19$\pm$0.05& 4384.25$\pm$43.80& 1.64$\pm$0.19& 5.6$\pm$0.3& $-$0.05$\pm$0.04\\
 658& 243.310654& $-$54.189079& 9.558& 1.705& $-$38.65$\pm$0.06& 4728.21$\pm$63.13& 2.04$\pm$0.17& 6.4$\pm$0.2& 0.07$\pm$0.03\\
 717& 243.337814& $-$54.165199& 9.502& 1.633& $-$39.31$\pm$0.14& 4731.75$\pm$51.29& 1.84$\pm$0.30& 12.7$\pm$0.4& $-$0.13+-0.05\\
 805& 243.352997& $-$54.334927& 9.317& 1.730& $-$37.86$\pm$0.06& 4443.43$\pm$33.71& 1.40$\pm$0.24& 4.7$\pm$0.4& $-$0.15$\pm$0.04\\
 895& 243.496063& $-$54.273777& 9.302& 1.902& $-$38.71$\pm$0.05& 4295.28$\pm$52.58& 1.04$\pm$0.16& 5.0$\pm$0.3& $-$0.10$\pm$0.05\\
 919& 243.433304& $-$54.281773& 9.747& 1.668& $-$39.03$\pm$0.07& 4994.09$\pm$42.96& 2.24$\pm$0.20& 7.8$\pm$0.2& 0.09$\pm$0.03\\
\hline
\end{tabular}
\end{center}
\end{table*} 

\subsection{Blue straggler stars}
The BSS are a distinct group of stars that are bluer and brighter than the MS turn-off in the CMD of the star cluster. The stars in an open cluster are considered as coeval populations, there should not be any star bluer than turn-off stars in the upper MS at the location of the BSS. The theories proposed to explain the straggler nature of the BSS stars have mostly one thing in common: the origin of the BSS is related to mass transfer to an MS star. The mass transfer could be from an evolving primary star caused by the Roche-lobe overflow or by collisions between single stars, binary, or triple stellar systems \citep{1964MNRAS.128..147M,1976ApL....17...87H}. We detected two BSS stars, which are shown in Figure~\ref{cmd_binary}. We estimated \textit{v}sin\textit{i} values for these BSS from the \textit{Gaia}-ESO spectra as 90$\pm$7 and 95$\pm$6 km s$^{-1}$. The relatively fast rotation of the BSS stars in NGC\,6067 compared to the general BSS populations seems to support the mechanism of formation through mass transfer in binary systems \citep{2023NatCo..14.2584F,2024ApJ...970..187N}.

\subsection{Be stars}\label{Be_sect}
\begin{figure}
	\includegraphics[width=8.5 cm]{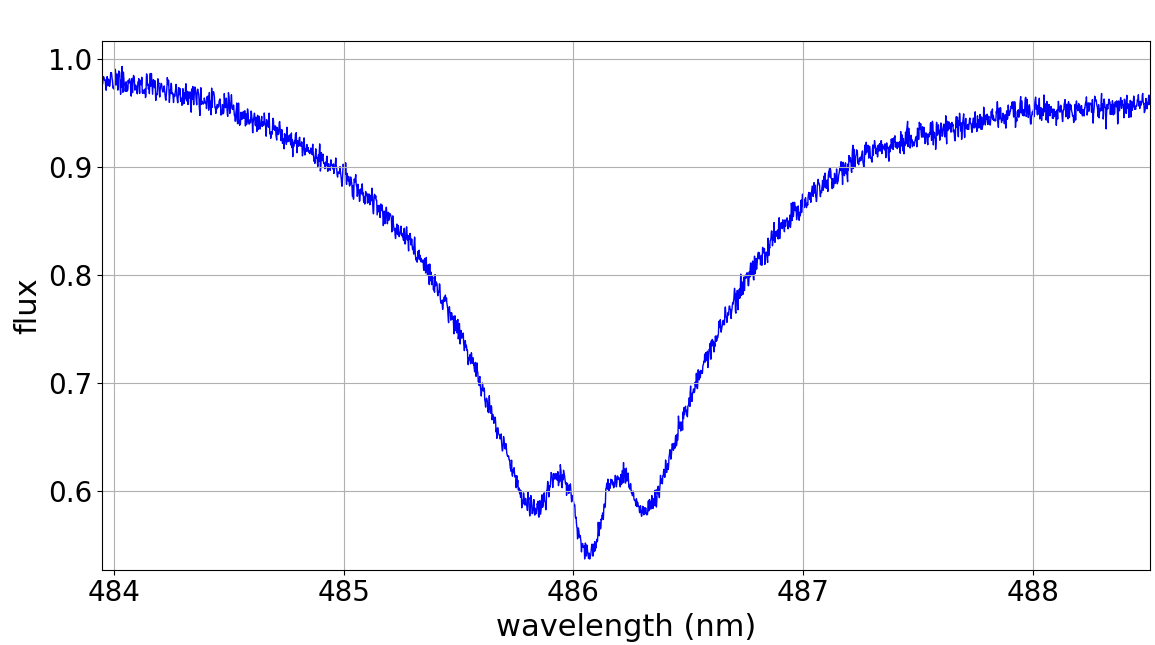}
        \includegraphics[width=8.5 cm]{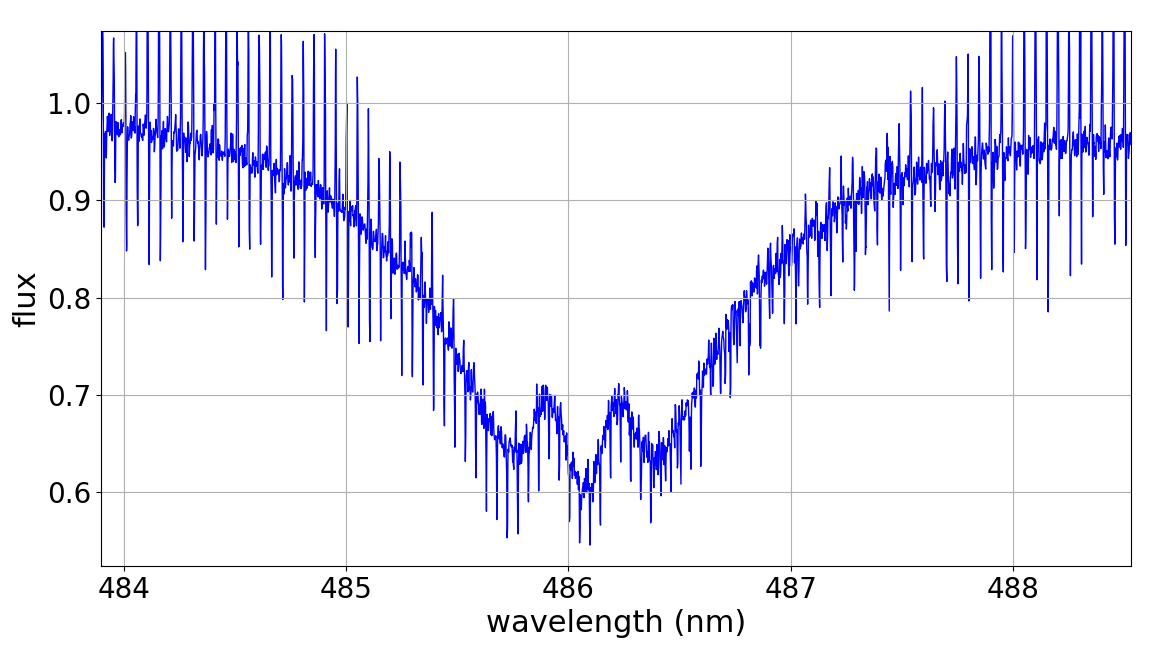}
    \caption{The normalized spectra of the Be stars with IDs 355 and 626 are shown in the upper and lower panels, respectively. The emission features are conspicuous in the H$_{\gamma}$ (486.1 nm) absorption line of these Be stars.}
    \label{Be_stars}
\end{figure}

The non-supergiant B-type stars with H$_{\alpha}$ emission are known as Be stars \citep{2003PASP..115.1153P}. The Be stars are very rapidly rotating stars hosting dust-free Keplerian disks \citep{2013A&ARv..21...69R}. The fast-rotating properties of the Be stars are useful in confirming the role of stellar rotation in producing the eMSTO in the clusters. We also identified two Be stars with IDs 355 and 626 having H$_{\beta}$ emission in the upper MS near the turn-off region of NGC\,6067. The spectra of these Be stars exhibiting emission features in H$_{\beta}$ absorption lines are shown in Figure~\ref{Be_stars}.
We estimated the \textit{v}sin\textit{i} values of these Be stars using the GIRFIT code. This code generates model spectra by interpolating in the grids of $\teff$ and $\logg$ values. The model spectra are further convolved to incorporate the rotational and the Gaussian instrumental broadenings. The GIRFIT code utilizes the SYNSPEC \citep{1995ApJ...439..875H} and ATLAS9 \citep{2003IAUS..210P.A20C} models to generate the synthetic spectra. We found \textit{v}sin\textit{i} values to be 155$\pm$15 and 197$\pm$25 km s$^{-1}$ for the Be stars with IDs 355 and 626, respectively. Furthermore, stars with IDs 402, 403, and 741 are also reported as Be stars by \citet{2017MNRAS.469.1330A}, but we lack spectra with a wavelength range containing H$_{\alpha}$ and H$_{\beta}$ for these stars to confirm their Be nature. However, we found large \textit{v}sin\textit{i} values of 295$\pm$26 and 326$\pm$20 km s$^{-1}$ for stars with IDs 402 and 741, which may be due to their Be nature. The Be stars found in NGC\,6067 are occupying the red part of the eMSTO in the CMD, which seems to support the stellar rotation scenario for the origin of the eMSTO in NGC\,6067, as Be stars are known to be rapidly rotating stars. The double peak in the emission feature of the H$_{\beta}$ line in the Be stars indicates them to belong to the shell stars class of Be stars (See Figure~\ref{Be_stars}). The shape of the emission profile depends on the inclination angle of the line-of-sight with the disk. The double peak emission feature corresponds to the equator-on view through the excretion disk surrounding the star \citep{2006A&A...459..137R}. 

\subsection{Spectroscopic Binaries}\label{SB2_sect}
\begin{figure}
        \includegraphics[width= 8 cm, height= 3.5 cm]{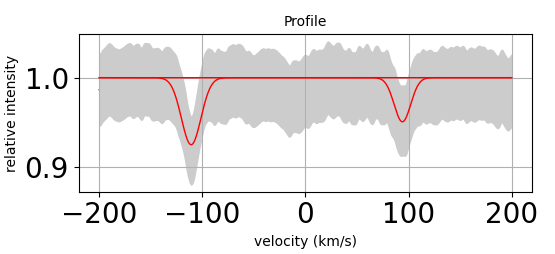}
        \includegraphics[width= 8 cm, height= 3.5 cm]{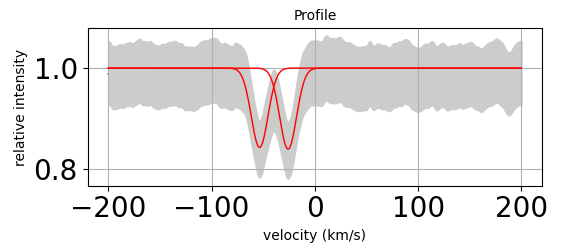}
	\includegraphics[width= 8 cm, height= 3.5 cm]{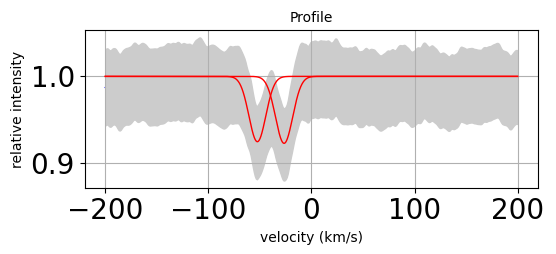}
        \includegraphics[width= 8 cm, height= 3.5 cm]{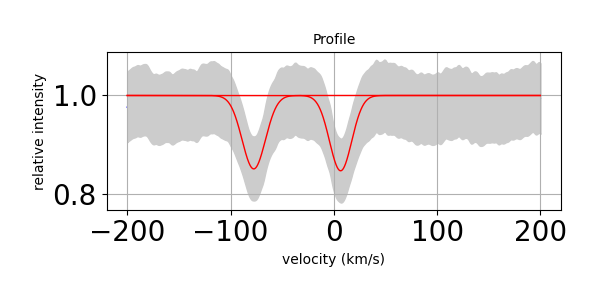}
    \caption{The cross-correlation function of SB2 stars with IDs 70, 609, 643, and 691 found in the eMSTO population of NGC\,6067 are shown from top to bottom order, respectively.}
    \label{RV_Binary}
\end{figure}

We found four spectroscopic binaries in the eMSTO population of the cluster NGC\,6067. These binaries were identified by double peaks in the cross-correlation function (CCF) and spectral line profiles \citep{2017A&A...608A..95M}. The star IDs of these binaries are 70, 609, 643, and 691. These binaries are the SB2, whose spectra contain fluxes from both the primary and secondary binary components. The CCFs of the SB2 binaries are shown in Figure~\ref{RV_Binary}. 
\begin{figure*}
    \includegraphics[width=17.5 cm]{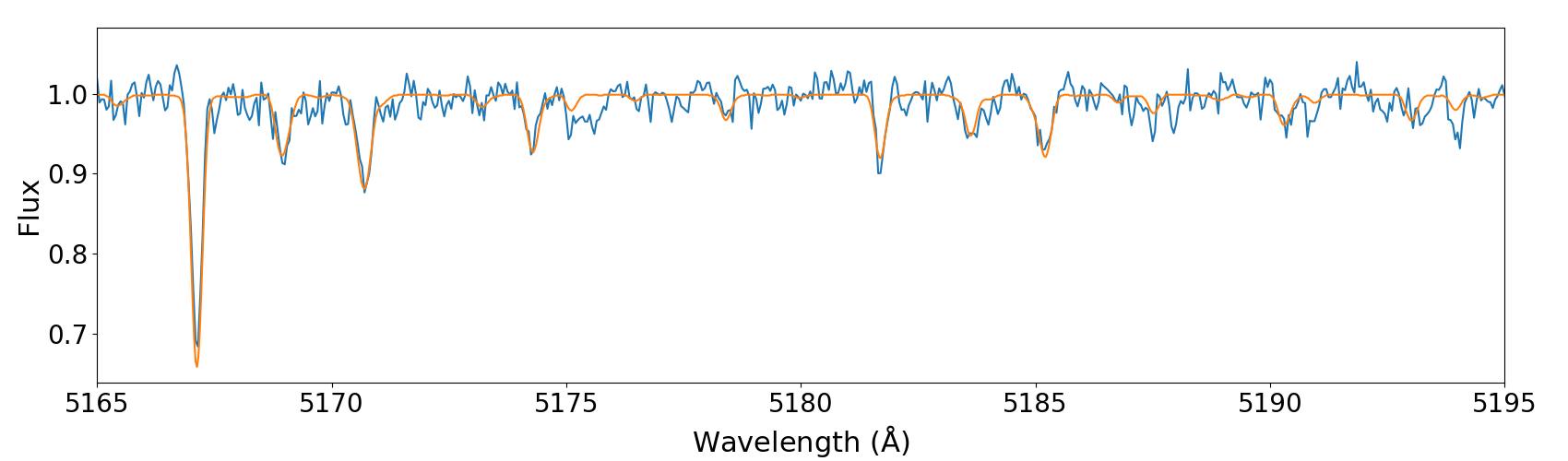}
	\includegraphics[width=17.5 cm]{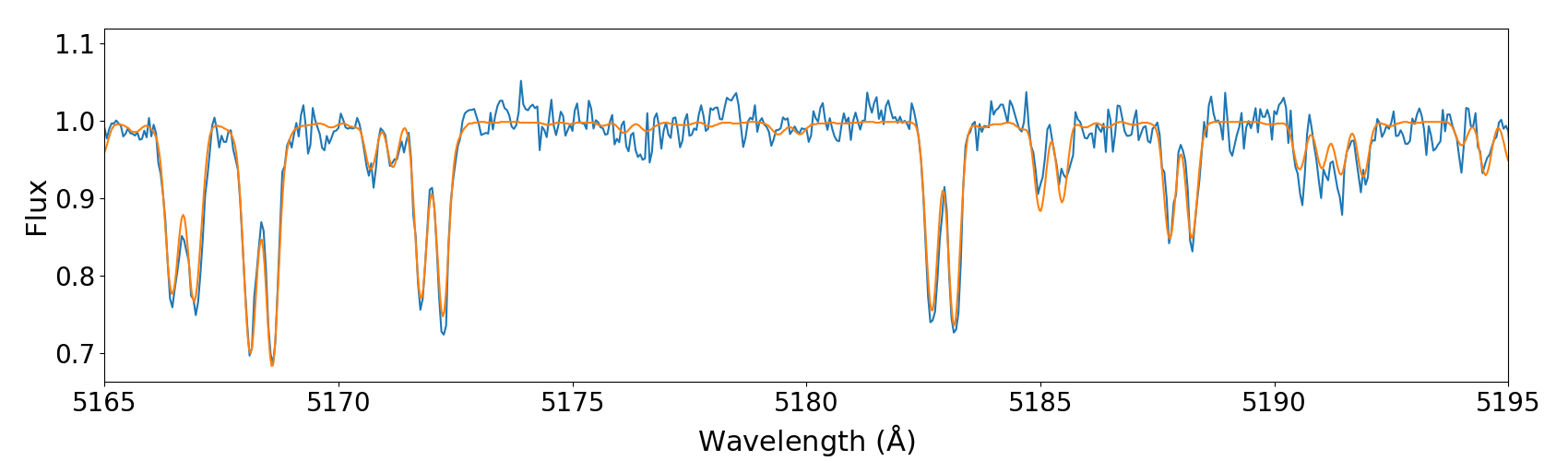}
    \includegraphics[width=17.5 cm]{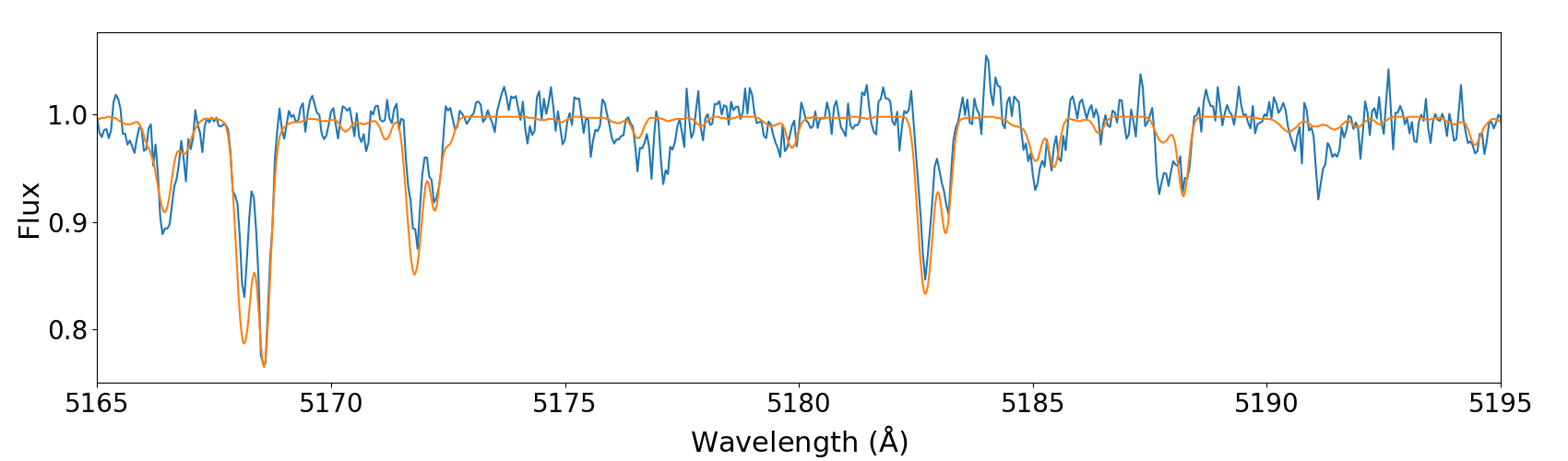}
    \includegraphics[width=17.5 cm]{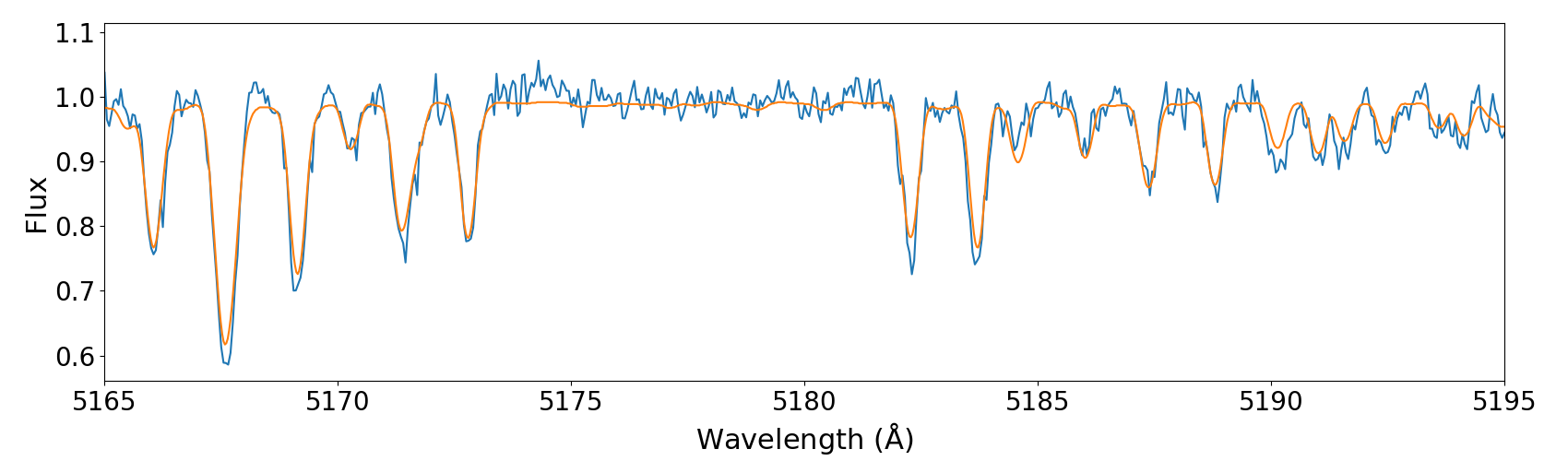}
    \caption{The spectra of the SB2 stars with IDs 70, 609, 643, and 691 are shown by the blue curves (top to bottom), illustrating the best-fit synthetic spectra expressed by yellow curves. The synthetic spectrum fitting for these binary stars is performed by the GIRFIT code.}
    \label{Binary_stars}
\end{figure*}   
The CCFs for these SB2 binaries are generated using the \textit{iSpec} package. We estimated \teff, \logg, \textit{v}sin\textit{i}, and RV values of the SB2 stars through a version of the GIRFIT code that takes into account stellar multiplicity \citep{2006A&A...451.1053F}. The code generates the synthetic spectra for each SB2 component separately and calculates their cross-correlation and corresponding radial velocity. The synthetic spectra are fitted to the observed spectra by minimization of the $\chi^{2}$, which is achieved through the MINUIT optimization package provided by the CERN. The MINUIT minimization package is based on the Nelder and Meade optimization algorithm. The best-fit spectra of the SB2 stars are illustrated in Figure~\ref{Binary_stars}. 

Unlike single stars in a cluster, these binary systems display distinct RV values, as their measured RV combine the cluster motion with the orbital velocities of the binary components. The \textit{v}sin\textit{i} values of these binary components are also the lowest among eMSTO stars possibly due to tidal interactions between binary components. The code is useful in estimating the velocity of the components and the atmospheric parameters of SB2 stars, especially when the components are blended. The estimated \teff, \logg, \textit{v}sin\textit{i}, and RV parameters of the SB2 components are given in Table~\ref{SB2_param}.     

\section{discussion}\label{discuss}
\subsection{Distribution of the \textit{v}sin\textit{i}}
The open cluster NGC\,6067 hosts various types of stars such as Be, SB2, BSS, RGB, and eMSTO stars. The rotational properties of these groups also exhibit very interesting distinctions. The two BSS stars are slow-rotating compared to the eMSTO stars of the cluster, with \textit{v}sin\textit{i} values of 90$\pm$7 and 95$\pm$6 km s$^{-1}$. In contrast, the RGB stars are much slower-rotating than the eMSTO stars with \textit{v}sin\textit{i} values below 20 km s$^{-1}$. Unlike the eMSTO stars, all the RGB stars are slow-rotating stars, as expected, due to the expansion of their envelope and further deepening of the convective envelope causing efficient magnetic braking \citep{2000ApJ...540..489S}.   
\begin{table*}
\caption{The physical parameters of the SB2 stars are estimated using the GIRFIT code. The IDs of the SB2 stars are given in column 1, and the luminosity ratios of the primary and secondary stars are given in column 10. The effective temperature, surface gravity, projected rotational velocity, and radial velocity of the primary components are given in columns 2, 3, 4, and 5, respectively, and those of the secondary components are listed in columns 6, 7, 8, and 9, respectively.} \label{SB2_param}
\begin{center}
\begin{tabular}{cccccccccc}  
\hline
    & \multicolumn{4}{c}{Component A}& \multicolumn{4}{c}{Component B} \\
    \cmidrule(lr){2-5} \cmidrule(lr){6-9}
ID  & \teff& \logg& \textit{v}sin\textit{i}& RV& \teff& \logg& \textit{v}sin\textit{i}& RV& lA/lB \\
\hline     
 70& 13103$\pm$89& 4.69$\pm$0.07& 8$\pm$0& $-$110.469$\pm$0.127& 9123$\pm$122& 3.92$\pm$0.24& 8$\pm$1& 93.023$\pm$0.368& 0.184$\pm$0.016 \\
 609& 8724$\pm$23& 3.04$\pm$0.05& 9$\pm$0& $-$53.315$\pm$0.087& 8860$\pm$22& 3.27$\pm$0.05& 8$\pm$0& $-$25.950$\pm$0.072& 1.051$\pm$0.022 \\
 643& 10946$\pm$46& 5.24$\pm$0.05& 13$\pm$0& $-$52.068$\pm$0.164& 9585$\pm$55& 1.81$\pm$0.08& 8$\pm$0& $-$26.854$\pm$0.109& 0.633$\pm$0.021\\
 691& 8412$\pm$26& 3.30$\pm$0.07& 14$\pm$0& $-$77.019$\pm$0.128& 8749$\pm$9& 3.35$\pm$0.06& 13$\pm$0& 5.852$\pm$0.128& 1.141$\pm$0.025\\
\hline
\end{tabular}
\end{center}
\end{table*}

\begin{figure*}
\includegraphics[width= 18 cm]{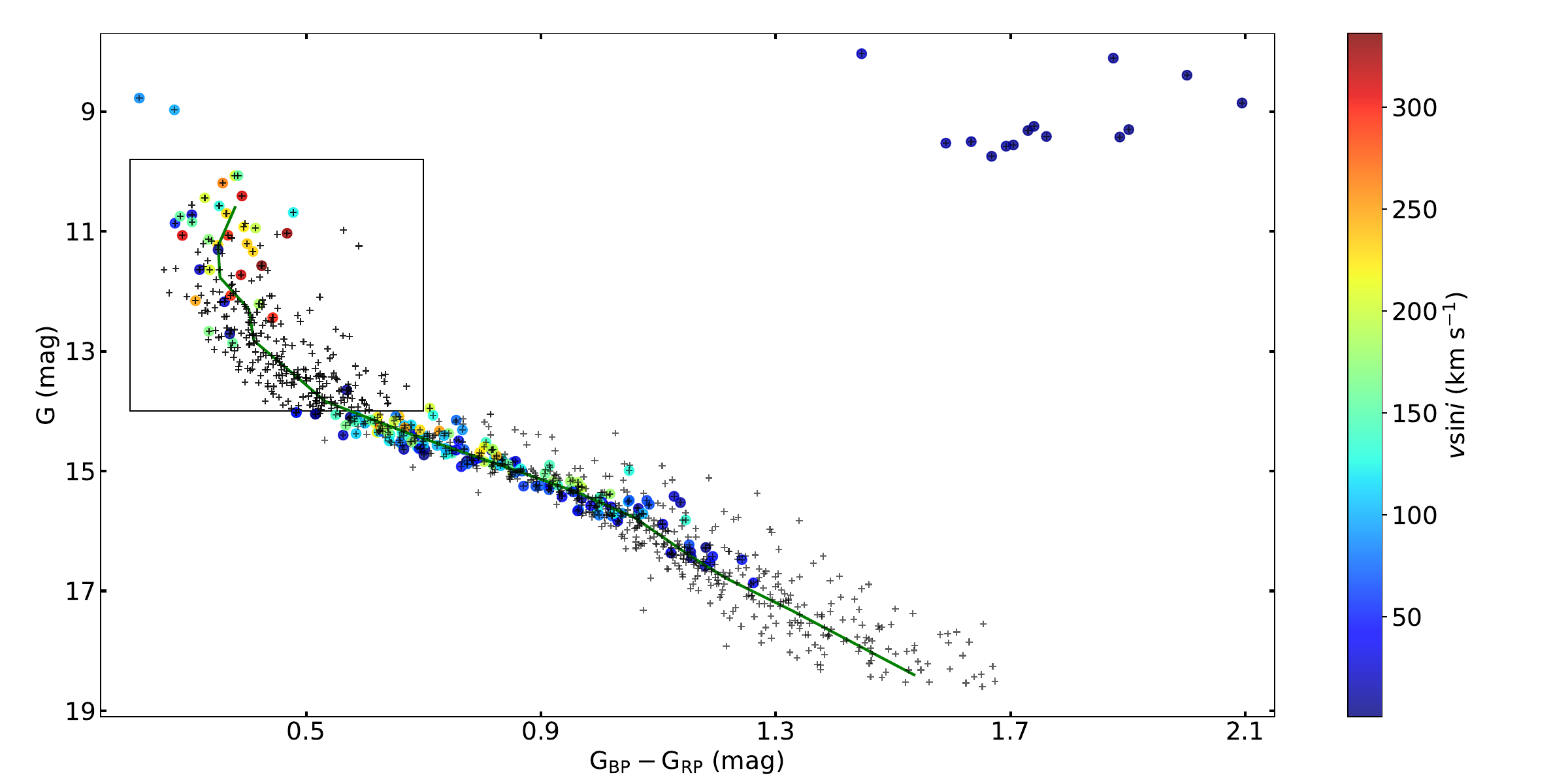}
    \caption{The color-magnitude diagram of NGC\,6067 color-coded by the projected rotational velocity of the stars. The black rectangle encloses the eMSTO stars of NGC\,6067. The green continuous curve represents the fiducial line for the main sequence stars.}
    \label{emsto}
\end{figure*}

\begin{figure}
\includegraphics[width=8.5 cm]{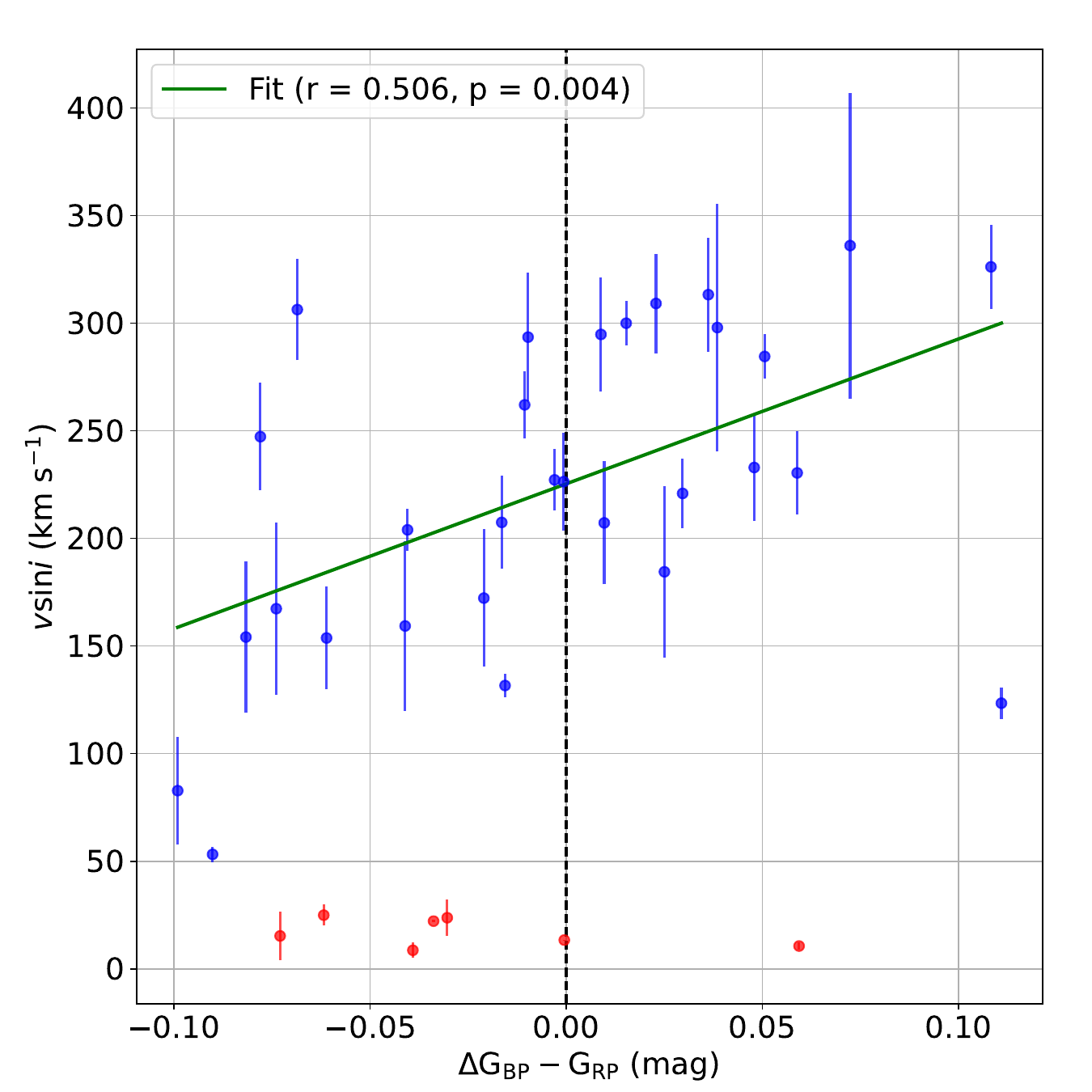}
    \caption{The correlation between pseudo-color $\Delta$G$_{\rm BP}-$G$_{\rm RP}$ and \textit{v}sin\textit{i} for the eMSTO stars. The red points show the outliers.}
    \label{correlation}
\end{figure}

Our main focus of the present study is the eMSTO stars. We found a trend that the stars in the red part of the upper MS are overwhelmingly fast-rotating. Slow-rotating stars are preferentially located in the blue part of the upper MS, as shown in Figure~\ref{emsto}. 
To further visualize the correlation between G$_{\rm BP}-$G$_{\rm RP}$ color and \textit{v}sin\textit{i} values for the eMSTO stars, we carried out a linear regression fitting to the data points as shown in Figure~\ref{correlation}. The pseudo-color $\Delta$G$_{\rm BP}-$G$_{\rm RP}$ shown on the x-axis of the plot is the color difference of the star with respect to the fiducial line. The fiducial line was created by interpolating the median G$_{\rm BP}-$G$_{\rm RP}$ color and G magnitude of the stars in magnitude bins of 0.5 mag. The points to the left of the fiducial line were assigned negative $\Delta$G$_{\rm BP}-$G$_{\rm RP}$ values, whereas those residing right of it were assigned positive values. 
The Pearson correlation coefficient for the fitting was found to be 0.506 with p-value = 0.004. We noticed a positive correlation between the pseudo-color $\Delta$G$_{\rm BP}-$G$_{\rm RP}$ and \textit{v}sin\textit{i} values of eMSTO stars, where redder stars generally rotate faster than their bluer counterparts, as also reported in previous studies \citep{2023MNRAS.518.1505K,2018ApJ...863L..33M,2019ApJ...883..182S}. The trend of the preferential location of the fast and slow rotating stars on the MS continues till G = 16.3 mag, starting from the turn-off point. This lower limit of the preferential locations corresponds to the mass of $\sim$1.6 M$_{\odot}$. 

The outliers excluded from the linear regression fitting are shown by red points in Figure~\ref{correlation}. The excluded points corresponds to eMSTO stars with IDs 21, 229, 317, 358, 403, 797, and 807 which have \textit{v}sin\textit{i} $<$ 50 km s$^{-1}$. A$_{V}$ values for three stars among these outliers with IDs 21, 229, and 797 are given to be 0.91$\pm$0.16, 0.84$\pm$0.20, and 0.91$\pm$0.22 mag, respectively by \citet{2024A&A...691A..98K}. These A$_{V}$ values are very close to the mean A$_{V}$ value of 0.93$\pm$0.01 for all the eMSTO stars in NGC\,6067. Thus, the outlier points for the correlation shown in the figure does not exhibit any significant extinction difference from the remaining eMSTO stars. These excluded slow-rotating stars could be binaries undetectable from the current set of spectroscopic data.

\begin{figure}
	\includegraphics[width=8.5 cm]{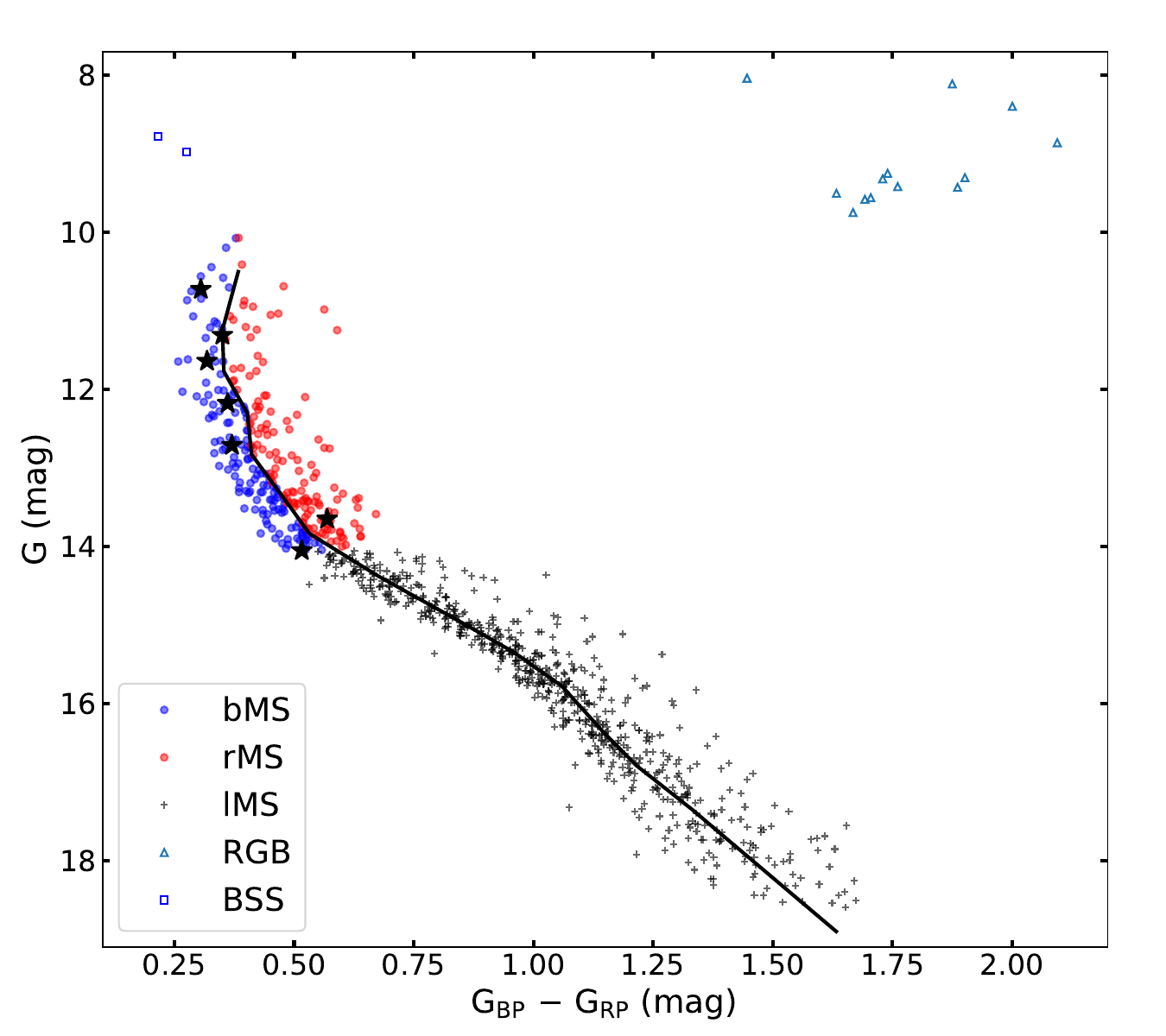}
    \caption{Plot exhibiting fiducial line (shown by the black line) and subgroups of stars in the color-magnitude diagram of NGC\,6067. The black asterisks represent outlier stars, as identified in Figure~\ref{correlation}.}
    \label{cmd_fiducial}
\end{figure}

We divided the MS stars into two groups for comparative study of \textit{v}sin\textit{i} distribution in them by drawing a fiducial line as shown in Figure~\ref{cmd_fiducial}. 
The bluer eMSTO stars than the fiducial line were grouped as bMS stars, and the redder stars were grouped as rMS stars. There were 146 bMS and 134 rMS stars in the eMSTO region of the cluster NGC\,6067. We could estimate the \textit{v}sin\textit{i} values for 23 bMS stars and 18 rMS stars depending on the availability of the spectra of the stars. The bMS stars have an average \textit{v}sin\textit{i} of 146$\pm$5 km s$^{-1}$, which are mostly slow-rotating stars. However, the rMS stars, excluding SB2 stars, are mostly fast-rotating stars having a mean \textit{v}sin\textit{i} value equal to 247$\pm$8 km s$^{-1}$. The spread in the distribution of the rotation rates of stars has also been previously found to be associated with the origin of the eMSTO phenomenon in the star clusters \citep{2019ApJ...883..182S,2024MNRAS.532.1212M}. However, the mean \textit{v}sin\textit{i} for groups of fast-rotating and slow-rotating stars for star clusters in the previous studies have significantly differed from their obtained values for NGC\,6067 \citep{2018ApJ...863L..33M,2023MNRAS.518.1505K,2024MNRAS.532.1212M}. The average \textit{v}sin\textit{i} value of a star also depends on its mass, as massive stars are expected to rotate faster. Hence, the younger clusters having eMSTO stars more massive than the older clusters are expected to have higher mean \textit{v}sin\textit{i} values for the bMS and rMS stars than older clusters. In our sample of eMSTO stars, we have spectra and thus \textit{v}sin\textit{i} values of generally brighter and more massive stars which may be a reason for the higher mean \textit{v}sin\textit{i} values for bMS and rMS stars than some previous studies. The metallicity of stars can also influence their rotation rates such that the rotation rates of the stars increase with decreasing metallicity \citep{2020ApJ...889..108A,2020MNRAS.499.3481A}. The star clusters in the Magellanic Clouds and the Milky Way have generally different metallicity, which may also cause different mean \textit{v}sin\textit{i} values for bMS and rMS stars. 

We found two Be stars and four spectroscopic SB2 binaries in the eMSTO population of the cluster NGC\,6067. The location of these Be stars is shown in Figure~\ref{cmd_binary}. The presence of the Be stars in the red part of the upper MS seems to further support the dependency of the color of the star on its rotational velocity, as the Be stars are known to be fast-rotating stars. The correlation between \textit{v}sin\textit{i} and G$_{\rm BP}-$G$_{\rm RP}$ color in NGC\,6067 could be due to gravity darkening. A few fast-rotating B-type stars may also have a redder color due to the dust-like extinction from their circumstellar excretion disk. 

The unresolved binary stars may also broaden the MS as they appear redder than the single stars on the MS. The spectroscopic binary stars are found to be located in the red part of the eMSTO region (see Figure~\ref{cmd_binary}). All the spectroscopic binaries detected in the eMSTO population of NGC\,6067 comprise slow-rotating components with \textit{v}sin\textit{i} values below 20 km s$^{-1}$. The slow rotation rates of the binaries might be due to tidal locking. We do not possess enough time-series spectra to efficiently detect the binary stars, especially the low mass-ratio binaries. Therefore, more binary stars might be present in the eMSTO population and we cannot completely rule out the contribution of the photometrically unresolved binaries in broadening the upper MS of NGC\,6067.   

The fast-rotating stars cease to exist below the mass $\sim$1.6 M$_{\odot}$ in NGC\,6067. The lower mass stars are expected to go through magnetic braking due to the strengthening of the magnetic field generated by their convective envelope. The stars go through a transition from the radiative envelope to the convective envelope near the mass of $\sim$1.6 M$_{\odot}$. The stars with convective envelopes developed stronger magnetic lines, causing the fast loss of angular momentum in the stars, known as the dynamo effect. The absence of the fast-rotating stars in the lower MS of NGC\,6067 below $\sim$1.6 M$_{\odot}$ seems to be caused by magnetic braking.   

\subsection{Tidal locking in binaries}\label{sec:tidal}
\begin{figure*}
	\includegraphics[width= 17.5 cm]{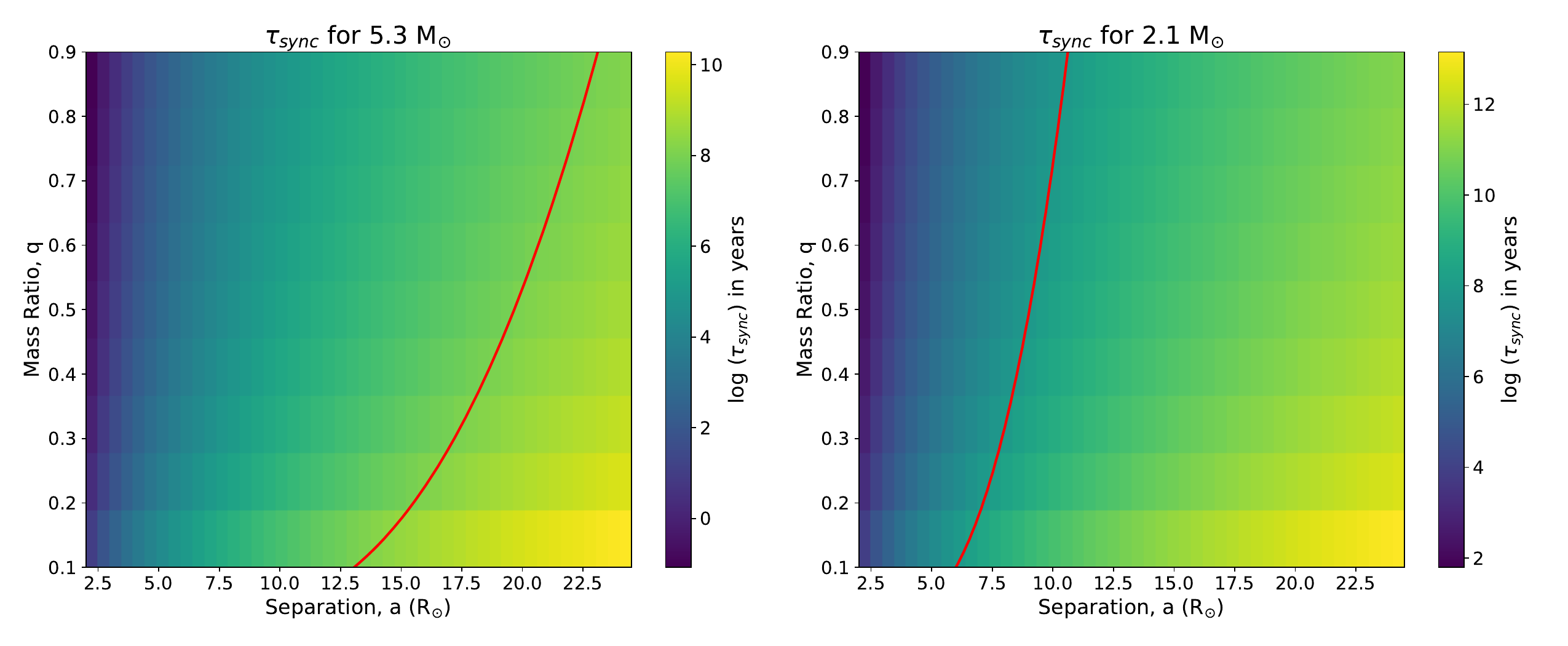}
    \caption{The plot for correlation between mass ratio, separation, and synchronization time for the binaries to be found in the eMSTO population of the cluster NGC\,6067. The colormap represents the synchronization time on a logarithmic scale. The red curve indicates the maximum separation for tidally synchronized binaries relative to the cluster age. The region to the left of the curve corresponds to synchronization times shorter than the cluster age, while the region to the right represents synchronization times longer than the cluster age. The masses corresponding to the synchronization time correlation plots are mentioned at the top of each subplot.}
    \label{sync_time}
\end{figure*}

The stars in the upper MS with approximately the same masses are expected to have a similar rotational velocity. However, the conspicuous spread in the distribution of the \textit{v}sin\textit{i} values of these stars encourages further exploration of the mechanism responsible for it. It has been suggested that all the stars were initially fast-rotating stars until tidal braking caused some of them to become slow-rotating \citep{2015MNRAS.453.2637D}. The initially fast-rotating binary systems change to the effectively tidal-locked binaries as their age reaches the synchronization time for the orbital motion. The synchronization time, $\tau_{sync}$, is the time required for a binary system to synchronize rotational and orbital motions due to tidal locking. We estimated the synchronization time using the following relations provided by \citet{2002MNRAS.329..897H}. 
\begin{center}
    $\frac{1}{\tau_{sync}}=5\times 2^{5 / 3}\left(\frac{G M}{R^3}\right)^{1 / 2} \frac{M R^2}{{I}} q^2\left(1+q\right)^{5 / 6} E_2\left(\frac{R}{a}\right)^{17 / 2}$
\end{center}
where $G$ denotes the gravitational constant. $a$ is the separation between primary and secondary. $I$, $M$, and $R$ symbolize the moment of inertia, mass, and radius of the primary component, respectively. A brief description for calculating the $\tau_{sync}$ is given in \citet{2024MNRAS.532.1212M}. We calculated the $\tau_{sync}$ values for the masses 2.1 and 5.3 M$_{\odot}$, representing the lower and upper bounds for the eMSTO stars in NGC\,6067. The $\tau_{sync}$ values for binary stars with these masses are shown in Figure~\ref{sync_time}. Since $\tau_{sync}$ for binary systems negatively correlates with the mass of the primary stars, the $\tau_{sync}$ for any binary present in the eMSTO sample would be between $\tau_{sync}$ values for 5.3 and 2.1 M$_{\odot}$ stars. We also have marked the maximum possible separation between binary components where $\tau_{sync}$ remains less than or equal to the cluster's age by a red curve in the figure.    
Based on the estimated synchronization time, we found that there could be close binaries whose $\tau_{sync}$ would be less than the age of the cluster NGC\,6067. Therefore, it is possible that some close binaries in the eMSTO of NGC\,6067 would be tidally locked. 
\begin{figure*}
	\includegraphics[width= 17.5 cm]{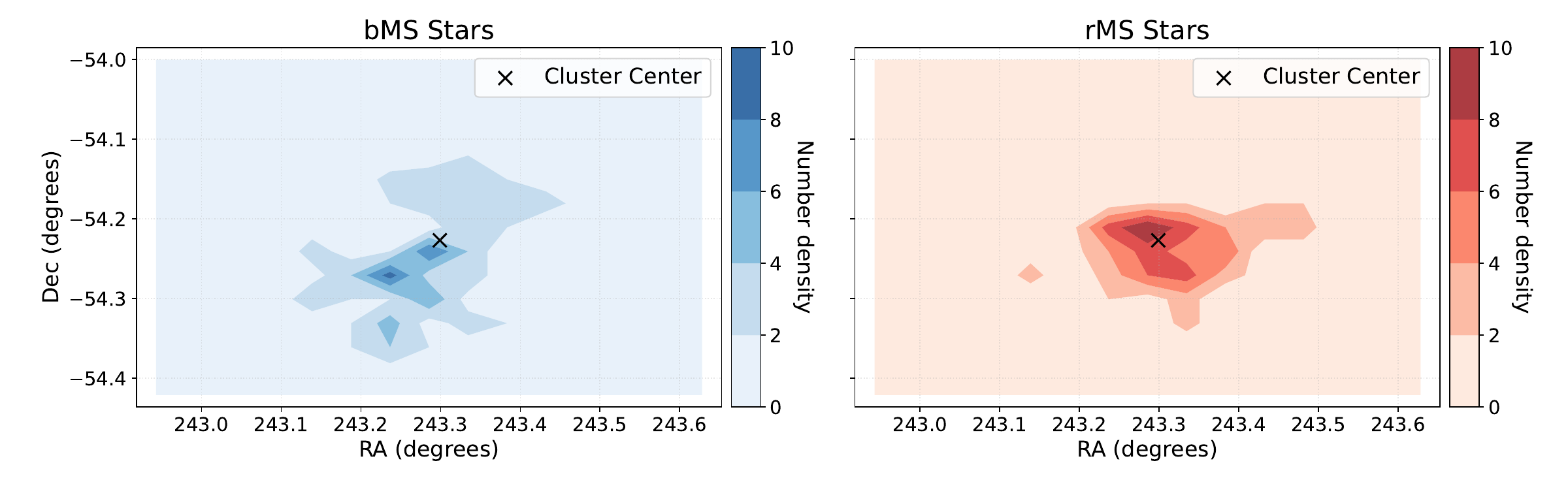}
    \caption{Spatial distributions of the bMS and rMS stars in the form of density contours in NGC\,6067. A black cross symbol marks the cluster center in the spatial distribution.}
    \label{spatial_emsto}
\end{figure*}
The role of the binaries in causing the slow-rotating stars in the blue part of the eMSTO can be assessed by examining the spatial distribution of the stars. If the slow-rotating eMSTO stars were tidally locked binaries then the bMS population, expected to consists predominantly of slow-rotating stars, should be preferentially located in the central region of the cluster due to mass segregation effect on these binaries. We calculated the dynamical relaxation time of the open cluster NGC\,6067 to investigate the possibility of mass segregation. Dynamical relaxation time is the timescale over which stars in a cluster approaches energy equipartition and a Maxwellian velocity distribution through two-body gravitational interactions among member stars. We calculated the dynamical relaxation time, T$_E$, using \citet{1971ApJ...164..399S} relation given below:
$$
T_E = \frac{8.9 \times 10^5 (N r_h^3/\bar{m})^{1/2}} {\log(0.4N)}
$$
where N denotes the total number of member stars. r$_{h}$ symbolize the half-mass radius of the cluster in parsec. The mean mass in solar mass unit of the member stars is denoted by $\bar{m}$. The half mass radius is the radial distance from the cluster center where cumulative mass becomes half of the total mass. The cluster center coordinates (RA = 243.29517 deg; Dec = -54.23182 deg) of NGC\,6067 were taken from \citet{2024A&A...686A..42H}. The masses of individual stars were estimated by fitting \citet{2017ApJ...835...77M} isochrone corresponding to the estimated age of the cluster on the CMD of NGC\,6067. We estimated $\bar{m}$ and r$_{h}$ to be 1.98 M$_{\odot}$ and 2.85 pc, respectively. We calculated T$_E$ to be 36.2 Myr using estimated $\bar{m}$ and r$_{h}$ values in the \citet{1971ApJ...164..399S} relation. The estimated T$_E$ value is less than cluster age of 91 Myr which suggest that cluster may exhibit mass segregation. We estimated mass segregation ratio (MSR) using edge length between stars in the minimum spanning tree (MST) method as suggested by \citet{2009MNRAS.395.1449A}. We used a Python package provided by \citet{Naidoo2019} to calculate the MST. The approach used to calculate the MSR is briefly described in \citet{2023JApA...44...71M}. This method is based on the basic principle that massive stars will have shorter mean edge length than that of the low-mass stars in the MST due to segregation. We obtained the MSR to be 1.2$\pm$0.1 for NGC\,6067. The MSR value very close to unity suggest a weak mass segregation in the cluster. Such a weak mass segregation may be due to the fact that the dynamical relaxation process has not yet fully completed. The obtained T$_E$ value of 36.2 Myr, significantly shorter than the cluster's age of 91 Myr, suggests that the cluster has likely reached some degree of dynamical relaxation. However, despite this, we observe only weak mass segregation in the cluster. One potential reason for this weak segregation could be that the dynamical relaxation time we calculated is likely an underestimate. This could be due to the fact that the number of detected member stars in NGC\,6067 might be smaller than the true number, owing to the completeness limit of the \textit{Gaia}-DR3 data. If the total number of member stars were higher, the relaxation time would be longer, implying that the cluster could still be in the process of dynamical evolution, rather than fully relaxed.   
\begin{figure}
	\includegraphics[width= 8.5 cm]{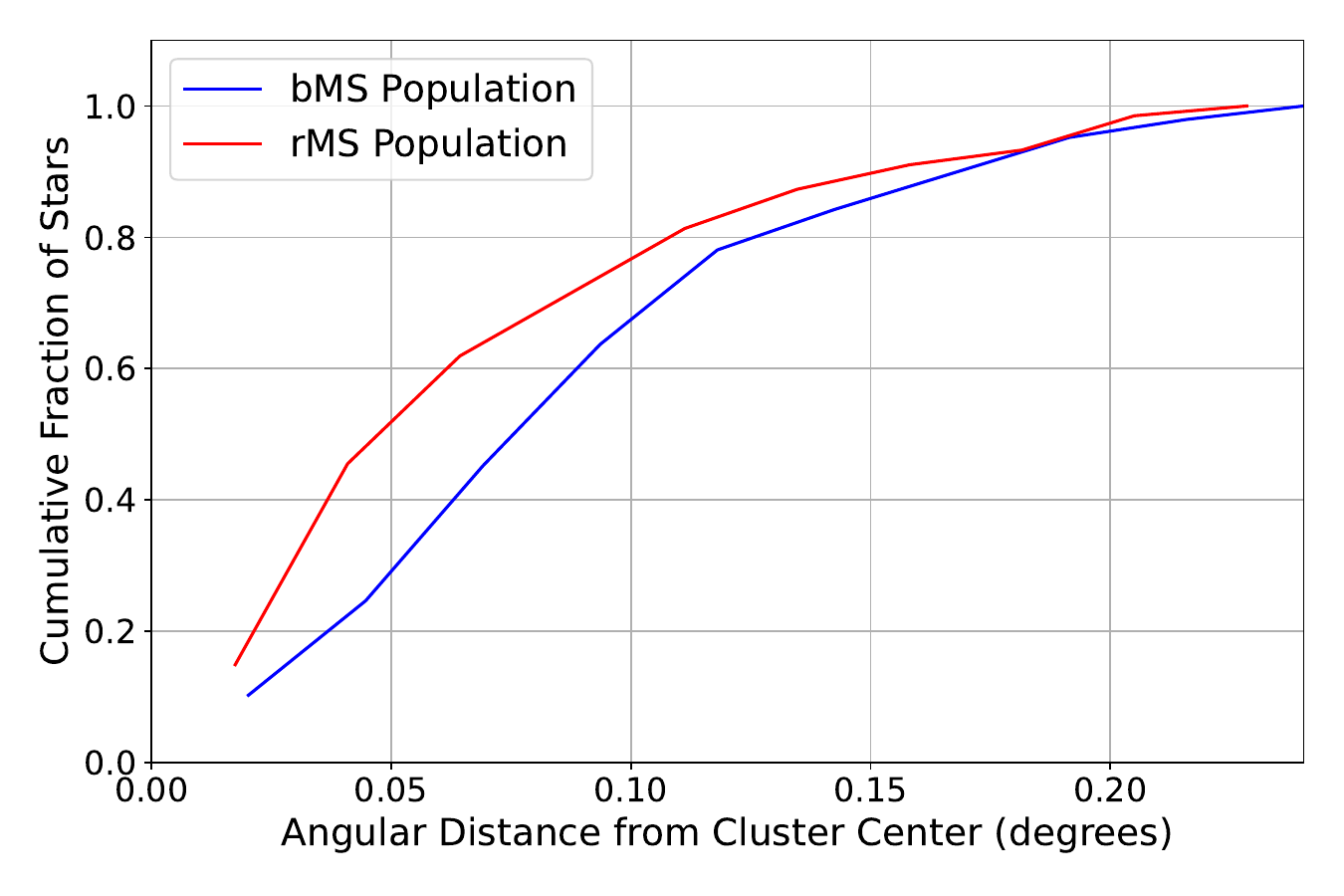}
    \caption{Cumulative radial distributions of the bMS and rMS stars in NGC\,6067.}
    \label{radial_emsto}
\end{figure}

The spatial distribution of the eMSTO stars in the form of density contours is shown in Figure~\ref{spatial_emsto}. For spatial distribution, we created a 15$\times$15 rectangular grid spanning the full range of RA and Dec coordinates. The star count in each cell was calculated as the number of stars falling within its boundaries, with empty cells assigned zero counts. We then generated smoothed density contours from these gridded counts to visualize the spatial patterns. The number density shown in the figure is the star count in each cell. 

A qualitative inspection of the spatial distribution of the eMSTO stars suggests that rMS stars are more centrally concentrated than bMS stars. This is further quantitatively confirmed by the cumulative radial distribution shown in Figure~\ref{radial_emsto}, where rMS stars exhibit a larger normalized cumulative fraction than the bMS stars up to an angular radial distance of 0.18 degrees. Notably, this angular radial distance is more than twice the core radius of 0.08 degrees obtained by \citet{2024A&A...686A..42H} for the cluster NGC\,6067. Beyond 0.18 degrees, the cumulative radial distribution of the bMS population closely follows that of the rMS stars. We tested the hypothesis that the bMS and rMS samples belong to the same radial distribution by performing the Kolmogorov–Smirnov (K-S) test on the samples. The K-S statistic for the bMS and rMS populations was 0.2609 with the p-value = 0.0001. The K-S statistics' obtained value indicates a notable difference in the radial distribution of the two populations. The p-value smaller than 0.05 supports the rejection of the null hypothesis that the two populations follow the same radial distribution. 

Although we find evidence of the weak mass segregation in the cluster, the bMS population (expected to comprise mostly tidally locked binaries) should be preferentially concentrated in the central region of the cluster if there were to be a preferential concentration of any population in the central region at all. The higher concentration of the rMS population, which hosts most of the fast rotating stars, in the central region of the cluster seems to indicate a factor other than the tidal locking in the binaries contributing to the spread of the rotation rates by slowing down a fraction of the eMSTO stars.   

\subsection{Environmental impact on stellar rotations}
The local environment in the PMS phase influences the rotation rates of the stars \citep{2021MNRAS.508.3710R}. The environmental influences include the role of far-ultraviolet radiation and dynamical interactions in the evolution and dissipation of the stars' protoplanetary disks. Stars may lose a substantial fraction of their angular momentum through the SDI during the first $\sim$10 Myr of their lifetime \citep{2014prpl.conf..433B}. The magnetized stellar winds may extract the angular momentum from the stars through open magnetic field lines, providing spin-down torque balancing the spin-up torque due to accretion on the star \citep{2005ApJ...632L.135M,2021ApJ...906....4I}. This spin equilibrium in the accreting star during the SDI phase is simplified as the 'disk-locking' process in the theoretical models. The star's rotation rate would be constant due to disk-locking during the lifetime of the disk. The disk-locking prevents the star from spinning up despite its contraction towards the zero-age MS during the PMS phase. After the destruction of the disk, stars become free to spin up while contracting in the PMS phase. Thus, the environmental influence on the disk dissipation time would also impact the rotational evolution of the stars.   

The longer SDI interval would lead to a slower rotation than the shorter SDI time, manifesting in the spread in rotation rates of the eMSTO stars in the cluster \citep{2020MNRAS.495.1978B}. The SDI mechanism could explain the observed spread in the rotation rates of the eMSTO stars in NGC\,6067. The intense radiation from the massive O/B-type stars in the central region of the open cluster causes faster disk dissipation through photoevaporation and, thus, a larger fraction of the fast-rotating stars \citep{2024AJ....167..120V}. This further explains the preferential concentration of the fast-rotating rMS stars in the central region of NGC\,6067 as visible in Figure~\ref{spatial_emsto}. The outskirts of the cluster NGC\,6067 preferentially host slow-rotating bMS stars, possibly due to longer disk dissipation time partly resulting from the lack of massive stars. Additionally, dynamical interactions in the central regions of the cluster can contribute to faster disk dissipation in the PMS phase of the central region stars due to higher density in the cluster center \citep{2021MNRAS.508.3710R,2024AJ....167..120V}.

\section{Conclusion}
We present a comprehensive analysis of the extended main sequence turn-off found in the open cluster NGC\,6067. We calculate the membership probability of the stars in the NGC\,6067 region and obtain a total of 944 member stars, including 280 member stars occupying the eMSTO of the cluster, using the HDBSCAN algorithm to detect the over-density in the proper motion plane. The age of the cluster is found to be 91 Myr through isochrone fitting on the color-magnitude diagram of the cluster. We utilize the medium-resolution spectra from the \textit{Gaia}-ESO archive available for 41 eMSTO (including four SB2 and two Be), 2 BSS, and 14 RGB stars to estimate the physical parameters of the individual member stars. We study the distribution of the \textit{v}sin\textit{i} for the eMSTO stars and the causes for the obtained distribution. The conclusions from the study can be summarized as follows:
\begin{enumerate}
    \item The bMS stars are mostly slow-rotating with a mean \textit{v}sin\textit{i} = 146$\pm$5 km s$^{-1}$ whereas predominately fast-rotating stars in the rMS population have the average \textit{v}sin\textit{i} = 247$\pm$8 km s$^{-1}$. We find a positive correlation between \textit{v}sin\textit{i} and the color of the eMSTO stars, which supports the hypothesis that the eMSTO can originate from the spread in rotation rates of the stars \citep{2009MNRAS.398L..11B}.

    \item We identify two Be stars from their spectra. The Be stars are situated in the right part of the tip of the MS turn-off, which strengthens the spread in rotation rates scenario for the origin of the eMSTO in NGC\,6067, as Be stars are known for their fast rotations. 

    \item The four SB2 binaries in NGC\,6067 occupy the red part of the eMSTO and comprise slow-rotating binary components. The locations of the SB2 binaries suggest that we can not completely rule out the contribution of the photometrically unresolved binaries in the observed broadening of the upper MS in the CMD of NGC\,6067.

    \item The synchronization time for the likely close binaries in the eMSTO population suggests that the close binaries in the eMSTO population would be tidally locked if present. However, the spatial distribution and the cumulative radial distribution of the eMSTO stars reveal a higher fraction of the fast-rotating rMS stars compared to the slow-rotating bMS stars in the central region of the cluster. This suggests tidal locking in the binaries to be a less likely mechanism for causing the spread in rotation rates of the eMSTO stars by slowing down a fraction of them.

    \item The SDI mechanism suggests that the stars retaining their protoplanetary disks longer tend to be slow rotators, while those with shorter disk lifetimes are typically fast rotators. According to the SDI theory, we would expect to find a higher concentration of fast-rotating stars in the central region of the cluster, as these stars experience more rapid disk dissipation due to photoevaporation and dynamical interactions in the cluster core. The spatial distribution and the cumulative radial distribution of the eMSTO stars seem to support this theory, indicating that the SDI mechanism is most likely responsible for the observed spread in rotation rates and the emergence of the eMSTO in NGC\,6067.  
    
\end{enumerate}

\section*{Acknowledgements} 

This research was supported by the Chinese Academy of Sciences (CAS) “Light of West China” Program (grant 2022-XBQNXZ-013), the Natural Science Foundation of Xinjiang Uygur Autonomous Region (grants 2022D01E86, 2023D01A12, and 2024D01B89), the Central Guidance for Local Science and Technology Development Fund (grant ZYYD2025QY27), the Tianshan Talent Training Program (grant 2023TSYCLJ0053) and the Tianchi Talent project. 

This work has made use of data from the European Space Agency (ESA) mission
{\it Gaia} (\url{https://www.cosmos.esa.int/Gaia}), processed by the {\it Gaia}
Data Processing and Analysis Consortium (DPAC,
\url{https://www.cosmos.esa.int/web/Gaia/dpac/consortium}). Funding for the DPAC has been provided by national institutions, in particular, the institutions participating in the {\it Gaia} Multilateral Agreement.

This research has made use of the tool provided by \textit{Gaia} DPAC (https://www.cosmos.esa.int/web/Gaia/dr3-software-tools) to reproduce (E)DR3 \textit{Gaia} photometric uncertainties described in the \textit{Gaia}-C5-TN-UB-JMC-031 technical note using data in \citet{2021A&A...649A...3R}.


\bibliography{manuscript}{}
\bibliographystyle{aasjournal}



\end{document}